\documentclass[floatfix,prx,superscriptaddress,twocolumn,letterpaper,nobalancelastpage]{revtex4-2}

\usepackage{mycommands}

\renewcommand{\section}[1]{\textit{#1}. ---}

\begin{document}
    
    \title{Dissipation Induced Flat Bands}
    
    \author{Spenser Talkington}
    \affiliation{Department of Physics and Astronomy, University of Pennsylvania, Philadelphia, Pennsylvania 19104, USA}
    \email{spenser@upenn.edu}
    
    \author{Martin Claassen}
    \affiliation{Department of Physics and Astronomy, University of Pennsylvania, Philadelphia, Pennsylvania 19104, USA}
    \email{claassen@upenn.edu}
    
    \date{\today}
    \begin{abstract}
        Flat bands are an ideal environment to realize unconventional electronic phases. Here, we show that fermionic systems with dissipation governed by a Bloch Lindbladian can realize dispersionless bands for sufficiently strong  coupling to an appropriately engineered bath. These flat bands emerge in a ``dark space'' of the system-environment coupling and are long-lived by virtue of symmetry protection from dissipation. We exhibit the robustness of this mechanism for general one and two band models with and without spin, and discuss conditions for their experimental realization such as in a 2D material on a superconducting substrate.
    \end{abstract}
    
    \maketitle
    
    \section{Introduction}\label{section:intro}
        Flat bands are a fascinating environment to realize unconventional quantum phases. With kinetic energy quenched, the behavior of electrons at partial filling of such bands is exclusively governed by Coulomb interactions and can lead to remarkable quantum collective behaviors, ranging from unconventional superconductivity to the fractional quantum Hall effect. While flat bands have long been appreciated to emerge from kinetic interference, as exemplified by the Kagom\'e \cite{syozi1951statistics} and Lieb lattices \cite{lieb1989two}, the discovery of energetically isolated almost-dispersionless bands in magic-angle twisted bilayer graphene \cite{suarez_morell_flat_2010,bistritzer_moire_2011} and other moir\'e heterostructures has garnered much recent attention as an experimentally accessible and highly tunable platform for flat bands. Very recent efforts have focused on cataloguing and classifying flat bands of crystalline materials \cite{regnault2021catalogue,calugaru21} and engineering the quantum geometry of flat bands \cite{julku2016geometric,liang2017band,herzog2022superfluid,bauer2022fractional,abouelkomsan2022quantum}.
        
        In this Letter, we present an alternate pathway to realizing flat bands that is independent of the underlying lattice structure and instead relies on the \textit{dissipative} coupling of a lower-dimensional quantum system---for instance a 2D material or 1D nanowire---to a higher-dimensional substrate, which acts as a bath. Central to our work, whereas the presence of a substrate is typically treated as an obstacle, we find that incoherent particle exchange with an appropriately engineered substrate can induce ``singular'' and spectrally isolated flat bands that counter-intuitively are long-lived in the limit of strong dissipation. Their origin can be traced to the emergence of a ``dark space'' of the system-bath coupling which is symmetry-protected from dissipation.
        
        We formulate a generic theory of such dissipation-induced long-lived flat bands for electronic systems. We then apply this theory to the most general spinless two-band model and illustrate our results for a paradigmatic example of a 2D topological insulator, and the most general inversion symmetric time-reversal symmetric one-band model with spin and illustrate our results for a triangular lattice, which could correspond to the isolated topmost band in a twisted bilayer transition metal dichalcogenide (TMD) such as WSe$_2$ \cite{wang2020correlated,devakul2021}.
        
        Over the last decade, combined dissipation and external driving has emerged as a way to drive systems into desired states \cite{diehl2011topology,bardyn2013topology,goldstein2019dissipation}.  Much emphasis has been placed on topologically classifying non-Hermitian Hamiltonians \cite{huang2014topological,lieu2018topological,gong2018topological,zhang2018topological}. Flat bands due to kinetic interference \cite{ramezani2017non}, and non-Hermitian particle hole symmetry \cite{ge2018non} have been studied in bosonic systems exhibiting classical gain and loss processes via non-Hermitian Bloch Hamiltonians.
        
        In quantum systems, a minimal model for dissipation that properly accounts for quantum jumps is provided by the Lindblad master equation $i \frac{\partial}{\partial t}\rho = \mathcal{L}[\rho]$ \cite{lindblad1976generators,gorini1976completely,gardiner2004quantum} for the reduced density matrix $\rho$, where
        \begin{align}\label{eq:GenericLindbladianCoherent}
    	    \mathcal{L}[\rho] = [H,\rho] - i \frac{\Gamma}{2} \sum_m \left(\{J_m^\dagger J_m,\rho\} - 2 J_m\rho J_m^\dagger\right)
    	\end{align}
    	describes the joint evolution under a coherent Hamiltonian $H$, together with dissipative processes governed via quantum jumps $J_m$ induced by the bath. These jump processes encode the fundamentally quantum processes of dissipation: measurement (collapse) processes and decoherence towards a thermal state over time---processes that are not captured by non-Hermitian Hamiltonians which exhibit the inherently classical phenomena of gain and loss. Here, the overall dissipation rate is parameterized by an energy scale $\Gamma$, and details of the quantum jumps are encoded in the form of operators $J_m$ with unit norm. This can be ``vectorized" so that $\hat{\mathcal{L}}\cdot\vec{\rho}=i\frac{\partial}{\partial t}\vec{\rho}$ in which case $\hat{\mathcal{L}}$ permits a matrix representation \cite{amshallem2015three,manzano2020short}.
    	
        In this Letter, we focus on the simple but conceptually rich scenario of a non-interacting system
        \begin{align}
            H(\bm{k}) = \sum_{\alpha,\beta} h_{\alpha\beta}(\bm{k}) \CD{\bm{k},\alpha} \C{\bm{k},\beta}
        \end{align}
        subjected to dissipative tunnel coupling to a bath. In their most general form, the corresponding quantum jump operators
        \begin{align}\label{eq:JumpOpLinear}
        	J_m(\bm{k}) = \sum_\alpha a_{m,\alpha}(\bm{k}) \C{\bm{k},\alpha} + b_{m,\alpha}(\bm{k}) \CD{-\bm{k},\alpha}
        \end{align}
        are linear in both electronic creation and annihilation operators, corresponding to particles tunneling in ($a_{m,\alpha}$) and out ($b_{m,\alpha}$) of the system. In particular, jump operators with $a\neq 0$ and $b\neq 0$ for the same $m$ are crucial to the emergence of a flat band as discussed below. These operators could arise from a superconducting or squeezed state bath \cite{altland2010condensed,walls1983squeezed,vaccaro1989phase}. The resulting master equation admits a tractable solution in terms of non-interacting fermionic (superoperator) normal modes \cite{prosen2008third,prosen2010spectral}, constituting a dissipative generalization of a band Hamiltonian to modes with finite lifetime. A recent work by Lieu, \textit{et al.} in Ref.~[\onlinecite{lieu2020tenfold}] classified the symmetry classes of such quadratic Lindbladians, and alternative classifications by entanglement eigenvalue crossings were proposed \cite{sayyad2021entanglement}.  Goldstein \textit{et al.} have considered using dissipation to drive systems into an almost flat band with non-trivial Chern number \cite{goldstein2019dissipation}, studied transport properties in these bands \cite{shavit2020topology}, and studied localization-delocalization transitions in these bands \cite{beck2021disorder}.
        
    \section{Formalism}
        We introduce a set of superoperator fermions that act on the left and right of the density matrix
        $
            \ell_{\k,\alpha} \dens = \C{\k,\alpha} \dens \Parity \text{ and } r_{\k,\alpha} \dens = \dens \CD{\k,\alpha} \Parity,
        $
        with fermion parity $\Parity = \exp(i\pi \sum_{\k,\alpha} \CD{\k,\alpha} \C{\k,\alpha})$. These operators can be viewed as a complex-fermion version of Prosen's third quantization algebra \cite{prosen2008third}, however they permit a treatment of even and odd parity sectors on equal footing. The operators obey the standard fermionic anticommutation relations $\{ \OP{\ell}{\k,\alpha}, \OPc{\ell}{\k',\alpha'} \} = \{ \OP{r}{\k,\alpha}, \OPc{r}{\k',\alpha'} \} = \delta_{\k,\k'} \delta_{\alpha,\alpha'}$ and $\{ \OP{\ell}{\k,\alpha}, \OP{\ell}{\k',\alpha'} \} = \{ \OP{\ell}{\k,\alpha}, \OP{r}{\k',\alpha'} \} = 0$. We note that this decomposition can be viewed as operators acting on forward and backward Keldysh contours \cite{altland2010condensed}.
        
        The Lindbladian can now be succinctly expressed as an operator $\hat{\mathcal{L}} = \bm{\Phi}^\dag \left[ L_{\textrm{coh}}(\k) - i L_{\textrm{dis}}(\k) \right] \bm{\Phi}$ in terms of fermionic fields $\bm{\Phi}_{\k} = ( \OP{\bm{\ell}}{\k},~ \OP{\bm{r}}{\k},~ \OPc{\bm{\ell}}{-\k},~ \OPc{\bm{r}}{-\k} )$ where boldface $\bm{\ell}$'s and $\bm{r}$'s denote vectors $\OP{\bm{\ell}}{\k} = (\OP{\ell}{\k,1}, \OP{\ell}{\k,1}, \dots, \OP{\ell}{\k,N})$ representing $N$ orbitals. The coherent and dissipative contributions take a $4N \times 4N$ single-particle matrix form $L_\text{coh}=\text{diag}(H_{\bm{k}},H_{\bm{k}},-H^\top_{-\bm{k}},-H^\top_{-\bm{k}})$ and
        \begingroup
            \setlength\arraycolsep{0.3pt}
        	\begin{align}
            L_\text{dis} \!=\! \frac{\Gamma}{2}
        	\!\!\begin{pmatrix}
        	A_{\k}\!-\!B_{\k}&-2B_{\k}&C_{\k}\!-\!C_{-\k}^\top&2C_{-\k}^\top\\
        	-2A_{\k}&B_{\k}\!-\!A_{\k}&-2C_{\k}&C_{\k}\!-\!C_{-\k}^\top\\
        	C_{\bm{k}}^\dagger \!\!-\!C_{-\k}^*&-2C_{-\k}^*&B_{-\k}^\top\!\!-\!A_{-\k}^\top & 2A_{-\k}^\top\\
        	2C_{\bm{k}}^\dagger&C_{\k}^\dagger \!\!-\!C_{-\k}^*&2B_{-\k}^\top& A_{-\k}^\top\!-\!B_{-\k}^\top
        	\end{pmatrix}
        	\end{align}
    	\endgroup
    	respectively, with $N \times N$ dimensional blocks
    	\begin{align}
    	     (A_{\k})_{\alpha,\beta} &= a_{m,\alpha}^*(\k) a_{m,\beta}(\k),\\ (B_{\k})_{\alpha,\beta} &= b_{m,\alpha}(-\k) b_{m,\beta}^*(-\k),\\ (C_{\k})_{\alpha,\beta} &= a_{m,\alpha}^*(\k) b_{m,\beta}(\k) 
    	\end{align}
    	determined by the jump operator amplitudes of Eq. (\ref{eq:JumpOpLinear}), and a sum over $m$ is implicit.
    	
    	The quadratic Lindbladian is of standard---albeit non-Hermitian---Bogoliubov de Gennes (BdG) form and can be diagonalized by a set of single-particle fermionic normal modes in analogy to a non-interacting Hamiltonian [see Supplemental Material]. Introducing a pseudospin representation for particles/holes and left/right contours in terms of Pauli matrices $\eta$ and $\tau$ makes the symmetries of the Lindbladian manifest. Analogous to a conventional BdG Hamiltonian, the Lindbladian is invariant under charge conjugation $(\bm{\ell}, \bm{r}) \to (\bm{\ell}^\dag, \bm{r}^\dag)$ represented by
    	$
            \mathcal{C}^{-1} L^\top(\k) \mathcal{C} = - L(-\k)
        $
        for $L = L_{\textrm{coh}}-i L_{\textrm{dis}}$ \footnote{The transposition for charge conjugation, as opposed to complex conjugation, stems from $L$ being non-Hermitian \cite{gong2018topological}.}, where
    	$\label{eq:ChargeConjugationSymmetry}
            \mathcal{C} = \eta_1 \otimes \tau_0 .
        $
        Hence, if $|\psi_R\rangle$ is a right eigenvector of $L$ with eigenvalue $\epsilon$, one can construct a left eigenvector $\langle \psi_L| = (|\psi_R\rangle)^\top \mathcal{C}$ with eigenvalue $-\epsilon$.
        
        Furthermore, $L$ obeys a ``contour reversal" symmetry
        \begin{align}
            \mathcal{T} = \eta_2 \otimes \tau_2,  \label{eq:ContourReversalSymmetry}
        \end{align}
        where
        $
            \mathcal{T}^{-1} [iL(\k)]^* \mathcal{T} = iL(-\k).
        $
        This defines a time-reversal (TR) like symmetry that squares to one and guarantees an eigenvalue $-\epsilon^*$ for each eigenvalue $\epsilon$ of $L$. Notably, as the quadratic Lindbladian accounts for both left and right contours propagating forward and backward in time, it \textit{always} exhibits a TR symmetry that exchanges contours even if the underlying Hamiltonian is not TR symmetric. Conversely, a TR symmetric system entails an \textit{additional} TR symmetry that acts only on a single contour (i.e. on $\OP{\ell}{}$, $\OP{r}{}$ fermions individually). Combined with charge conjugation, this implies that all eigenvalues of $L$ come in quadruplets $\pm \text{Re}(\epsilon)\pm i\,\text{Im}(\epsilon)$, with the particle-like modes (with negative imaginary part) representing the physical excitation spectrum with finite lifetimes $\hbar/|\text{Im}(\epsilon)|$. Finally, the combination of charge conjugation and TR forms a chiral symmetry $\mathcal{S}^{-1} [iL(\k)]^\dag \mathcal{S} = -iL(\k)$ with $\mathcal{S} = i \eta_3 \otimes \tau_2$.

    \section{Dissipative Dark Space}
        Remarkably, imposing one additional unitary symmetry
        $
        \mathcal{D} = \eta_3\otimes\tau_1,
        $
        that commutes with $L$ can now be shown to dictate the emergence of a dissipationless ``dark space''---the dissipationless subspace of the purely dissipative Lindbladian that describes coupling to the \textit{bath}---which guarantees the persistence of long-lived bands of the \textit{system} even in the limit of strong system-bath coupling. Exceeding a critical system-bath coupling strength necessitates the formation of a long-lived flat band. We will see that this symmetry is naturally satisfied for coupling to a superconducting substrate. Whereas the coherent contribution trivially commutes with $\mathcal{D}$ from its pseudospin representation \footnote{Here we show the case where $H:=H_{\bm{k}}=H_{-\bm{k}}$. For the general case, see the [Supplemental Material].}
        \begin{align}
    	    L_\text{coh} = \Re(H)\eta_3\otimes\tau_0+i\,\Im(H)\eta_0\otimes\tau_0,
    	\end{align}
    	$\mathcal{D}$ imposes strong constraints on the form of the system-bath coupling by demanding that $A_{\k} = B_{\k}$ and $C_{\k} = C_{-\k}^\top$. An ansatz that fulfills this symmetry is
        \begin{align}\label{eq:ansatz}
        (b_{m,1},\dots,b_{m,N})^\top = e^{iS}(a_{m,1},\dots,a_{m,N})^\top,
        \end{align}
        where $S$ is any real, symmetric matrix \footnote{We conjecture that this is the \textit{only} form that ensures $\mathcal{D}$ symmetry.}. In this case, the dissipative part of the Lindbladian can be written as
        \begin{align}\begin{split}
    	    L_\text{dis} = -\Gamma\bigg(&\Re\left(A\right)\eta_3\otimes\tau_1 +i\,\Im\left(A\right)\eta_0\otimes\tau_1\\
    	    +&\Im\left(C\right)\eta_1\otimes\tau_2+\Re\left(C\right)\eta_2\otimes\tau_2\bigg).
    	\end{split}\end{align}
    	for the simplest inversion symmetric case where $A=A_{\bm{k}}=A_{-\bm{k}}$ and $C=C_{\bm{k}}=C_{-\bm{k}}$; for the general case see the [Supplemental Material].
    	Remarkably, one finds that $L_\textrm{dis}$ becomes a \textit{Hermitian} operator due to the dark space symmetry constraint.
    	
    	Invariance under $\mathcal{D}$ has important consequences for the eigenspectrum of the dissipative part of the Lindbladian. The latter can be shown using a unitary rotation $U = \textrm{diag}( 1,-1,1,1 )$ to decompose as
    	\begin{align}
    	    U L_{\textrm{dis}} U^\dagger = -\Gamma\, \tau_1 \otimes
    	    \begin{pmatrix}
    	    \bm{a}^\dagger(\k)\bm{a}^{\vphantom{\dagger}}(\k) & \bm{a}^\dagger(\k)\bm{b}^{\vphantom{\dagger}}(\k)\\
    	    \bm{b}^\dagger(\k)\bm{a}^{\vphantom{\dagger}}(\k) & \bm{b}^\dagger(\k)\bm{b}^{\vphantom{\dagger}}(\k)
    	    \end{pmatrix}, \label{eq:dissipativeDecomposition}
    	\end{align}
    	where $[\bm{a}]_{m,\alpha} = a_{m,\alpha}$, $[\bm{b}]_{m,\alpha} = b_{m,\alpha}$ are $N \times N$ matrices that parameterize the jump operators, and we used the fact that $\mathcal{D}$ symmetry guarantees $[\bm{a}^\dag(-\k) \bm{a}(-\k)]^* = \bm{b}^\dag(\k) \bm{b}(\k)$ and $[\bm{a}^\dag(\k) \bm{b}(\k)]^* = \bm{b}^\dag(\k) \bm{a}(\k)$.
    	
    	The tensor decomposition illustrates that each eigenmode of the right-side $(2N\times 2N)$ operator with energy $\epsilon$ comes with a charge-conjugate mode with energy $-\epsilon$, as expected. More importantly however, the eigenspectrum of the right-side operator guarantees, for an $N$-band system, there are exactly $N$ zero modes. These zero modes span a ``dark space'' that is protected from dissipation; explicitly, they read
    	$
    	    |\phi_i^\pm\rangle = U^\dag \left(\bm{u}_i,\, \bm{v}_i,\, \pm \bm{u}_i,\, \pm \bm{v}_i \right),
    	$
    	where $\pm$ indexes particle-like and hole-like modes, and $\bm{u}_i$, $\bm{v}_i$ are $N$-dimensional vectors in orbital space that compose the $i^\textrm{th}$ solution of the equation
    	$
    	    \bm{a} \bm{u}_i + \bm{b} \bm{v}_i = 0,
    	$
    	and hence depend on details of the jump operators.
    	The remaining eigenmodes of $L_{\textrm{dis}}$ have a finite lifetimes proportional to $1/\Gamma$. However, further zero modes are possible if the number of jump operators is $M<N$; then $\bm{a}$ and $\bm{b}$ have rank $M$, guaranteeing an additional set of $N-M$ zero modes, similar to a Hermitian case discussed in Ref. [\onlinecite{hatsugai2011zq}-\onlinecite{mizoguchi2019molecular}]. Overall, one obtains $2N-M$ particle-like zero modes, and an equal number of hole-like zero modes.
    	
    \section{Flat Bands from Dissipation}
    	We now turn to implications of this ``dark space'' on the spectrum of the full Lindbladian $L_{\textrm{coh}} - i L_{\textrm{dis}}$. Let the dark space be spanned by a set of particle-like zero modes $| \phi_i \rangle$ of $L_{\textrm{dis}}$, and suppose that the overall dissipative coupling $\Gamma$ is large with respect to the coherent energy scales of the band Hamiltonian. In this limit, $L_{\textrm{coh}}$ can be treated as a ``coherent'' perturbation on $L_{\textrm{dis}}$. The lowest-order contribution follows from projecting $L_{\textrm{coh}}$ into this (at least $N$-dimensional) dark space. Diagonalizing the resulting effective Lindbladian $\widetilde{L}_{ij} = \langle \phi_i | L_{\textrm{coh}} | \phi_j \rangle$ yields $N$ bands with infinite lifetime, which are generically dispersive.
    	
    	However, contour reversal symmetry $\mathcal{T}$ [Eq. (\ref{eq:ContourReversalSymmetry})], which commutes with dark space symmetry and obeys $\mathcal{T}^2 = 1$, imposes a strong constraint, as each band $\epsilon(\bm{k})$ projected into the dark space must come with a conjugate partner $-\epsilon^*(-\bm{k})$. If the system additionally obeys inversion symmetry $\mathcal{I}$, this guarantees a partner state $-\epsilon^*(\bm{k})$ for each state $\epsilon(\bm{k})$; if the dimension of the dissipative dark space (which in general equals the number of bands, as argued above) is \textit{odd}, contour reversal symmetry thus necessitates a single ``dangling'' mode at zero energy $\epsilon=0$ that remains invariant under $\mathcal{T} \mathcal{I}$. This mode hence forms a stable flat band for all Bloch momenta.
    	
    	Second-order corrections in $L_{\textrm{coh}}$ attribute a finite lifetime to this dissipation-induced flatband. As such processes scale as $1/\Gamma$, a strong dissipative coupling $\Gamma$ counter-intuitively ensures a long lifetime $\tau \propto \Gamma$ for the flat band \footnote{Here, factors proportional to the electronic bandwidth have been omitted for brevity.}. This is because second-order corrections necessarily involve an intermediate state with short lifetime $1/\Gamma$ that couples the dissipationless dark space to the dissipative subspace so the two subspaces decouple in the strong dissipation limit.
    
        The above mechanism readily generalizes to account for electron spin. Here, each contour is \textit{individually} invariant under spinful time-reversal symmetry (TRS) $\mathcal{T}_{\textrm{trs}}$ with $\mathcal{T}^2_{\textrm{trs}} = -1$. Crucially, $\mathcal{T}_{\textrm{trs}}$ commutes with contour reversal $\mathcal{T}$. Modes that span the dissipative dark space now come in Kramer's pairs $\epsilon_\nu(\k), \epsilon_{\bar{\nu}}(-\k)$, where $\nu, \bar{\nu}$ denote spin-orbital indices; if the system again also obeys inversion $\mathcal{I}$, the dark space splits into two degenerate subsectors of opposite spin for each $\k$, analogous to spin-degenerate bands in a centrosymmetric system with TRS. For each spin sector, the same rules of zero-mode counting apply, and a stable flat band again forms for all Bloch momenta if the number of bands (or more precisely, the dimension of the dark space) \textit{per spin} is odd.
    
    \section{Example: Generic Spinless Two-Band Model}
        \begin{figure*}
	        \includegraphics[width=1.93 in]{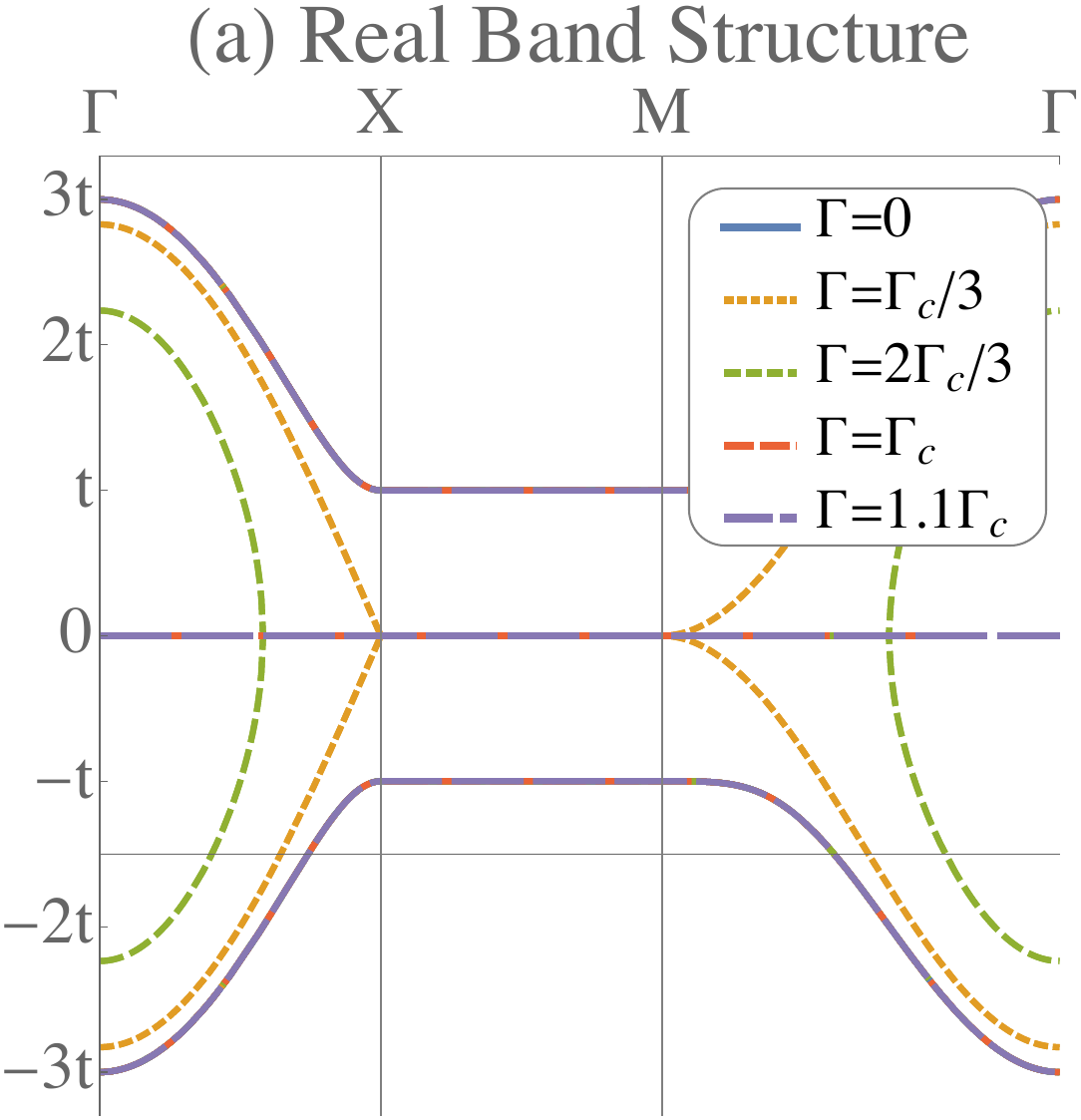}\ \ \ \
	        \includegraphics[width=1.93 in]{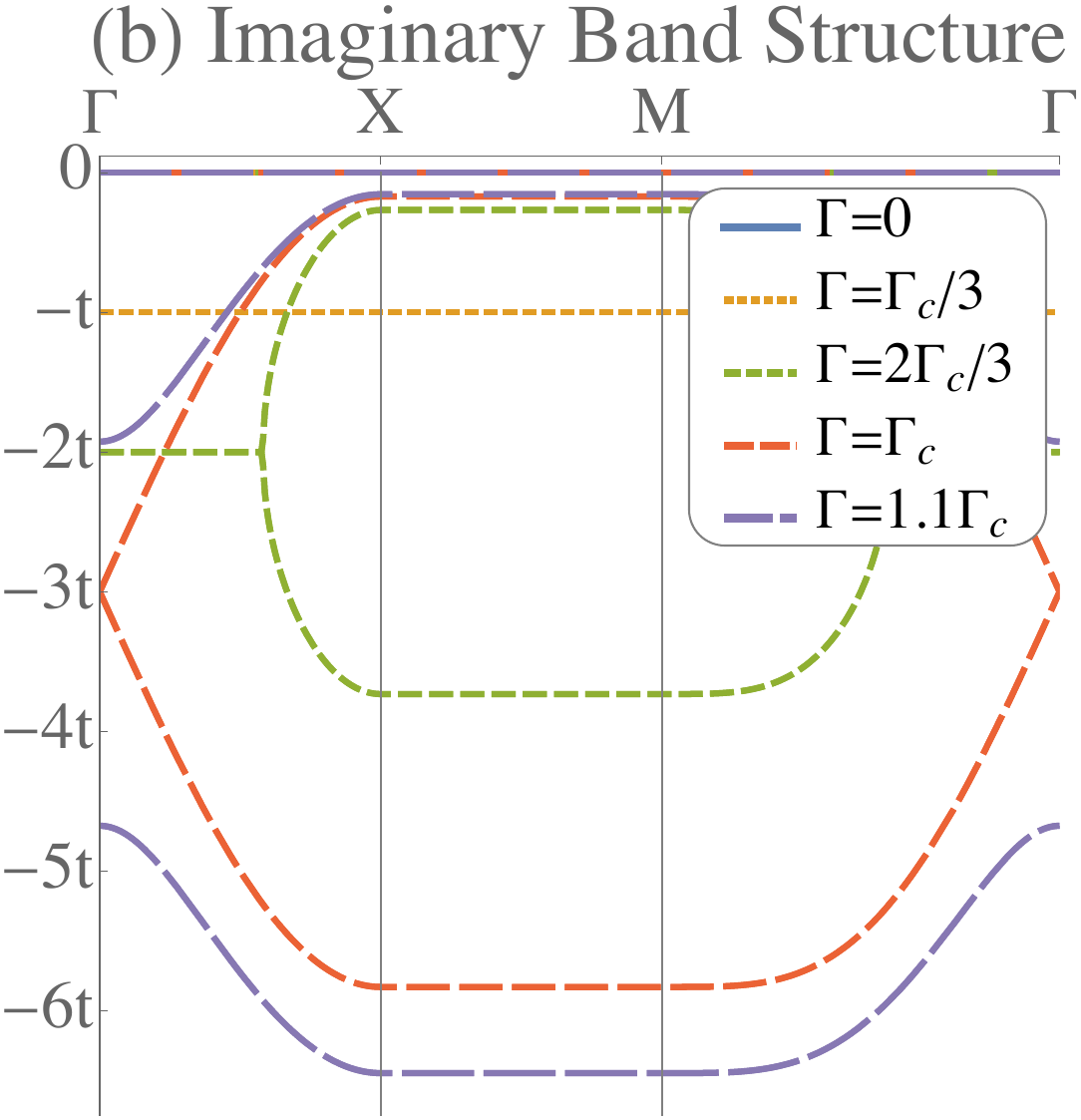}\ \ \ \
	        \includegraphics[width=2.85 in]{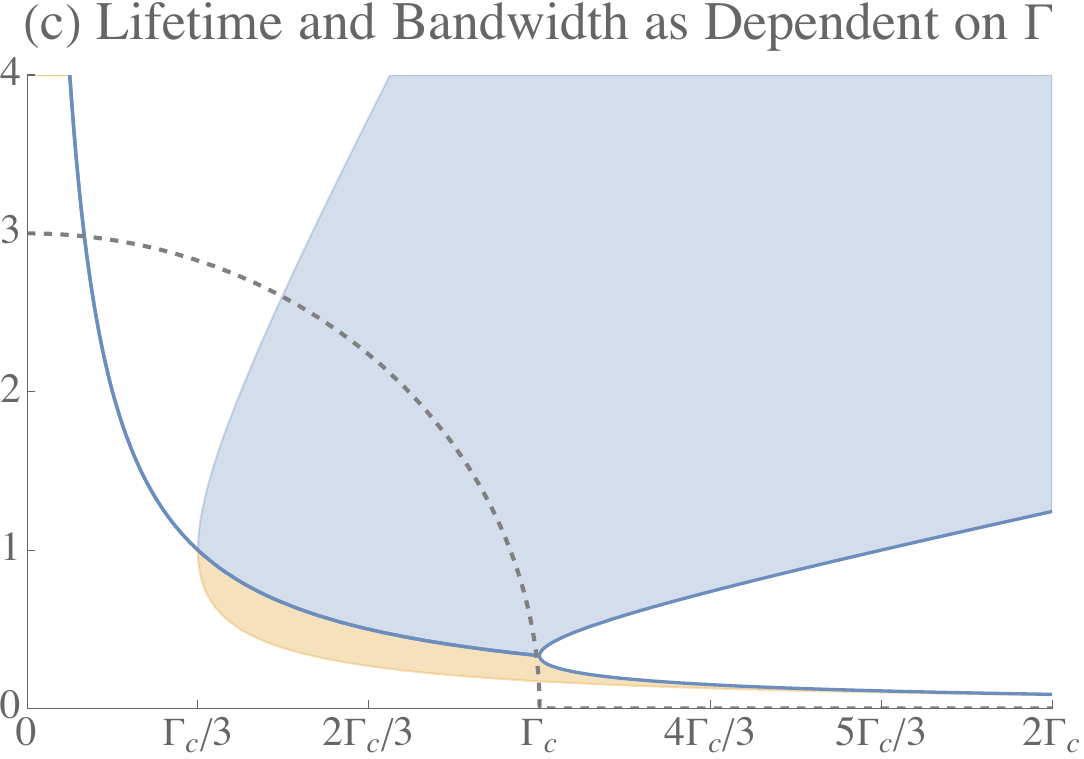}
	        \caption{\label{fig:qwz} \textbf{Band structures and lifetimes of the Qi-Wu-Zhang model dissipatively coupled to a bath} ($m=1$) for a variety of dissipation rates. \textbf{(a)} In the zero dissipation limit, only dispersive bands are present. The bandwidth of the bands coupled to the bath narrows with increasing dissipation. \textbf{(b)} For zero dissipation, all bands have zero imaginary part and are long-lived, while for finite dissipation, bands coupled to the bath acquire an imaginary part and finite lifetimes. In the strong dissipation limit, there are ``long-lived" bands whose imaginary parts tend to zero, and ``short-lived" bands whose imaginary parts tend to negative infinity. \textbf{(c)} Lifetime of the dissipative states as a function of dissipation (units $\hbar/t$). The solid lines are at $(k_x,k_y)=(0,0)$, the blue region encapsulates the lifetimes of the ``long-lived" flat bands, whose lifetimes increases linearly with $\Gamma$ for large $\Gamma$, and the orange region encapsulates the lifetimes of the ``short-lived" flat bands, whose lifetimes decreases as $1/\Gamma$ for large $\Gamma$. The dashed gray line is the bandwidth (units $t$) of the dissipative bands which is zero for $\Gamma\geq\Gamma_c$.}
	    \end{figure*}
            As a first candidate, consider a generic two-band Hamiltonian 
    	    $
    	    H(\bm{k}) = \vec{d}(\bm{k})\cdot\vec{\sigma} 
    	    $
    	    where $\sigma_i$ are the Pauli matrices.
            Now, the simplest form of dissipation that satisfies Eq.~(\ref{eq:ansatz}), is 
    	    \begin{align}
    	    J = c_{\bm{k},m}+c_{\bm{k},m}^\dagger,  \label{eq:jumpSimple}
    	    \end{align}
    	    with $m=1$ \textit{or} $2$. This jump operator describes the creation/annihilation of a Majorana fermion in the electronic system. As the number of bands is even, we choose a single jump operator to ensure an odd-dimensional dark space via rank-deficiency. A natural physical realization would be a system with an in-plane and out-of-plane orbital, with only the latter coupled to the substrate.
    	    
    	    The resulting single-particle modes of $L$ can be found exactly and break into two groups: the modes of the dissipationless system, $\epsilon_{\pm}^s = \pm |\vec{d}|$, and
    	    \begin{align}\label{eq:d.sigma-modes}
    	    \epsilon_{\pm\pm}^u = \pm \Gamma\sqrt{|\vec{d}|^2/\Gamma^2-2 \pm 2\sqrt{1-|\vec{d}|^2/\Gamma^2}}\ ,
    	    \end{align}
    	    The $\epsilon^s$ states are stable and the $\epsilon^u$ states acquire a finite lifetime with $\Gamma>0$. 
    	    
    	    As an example, we consider the Qi-Wu-Zhang model of a two-dimensional Chern insulator \cite{qi2006topological}, for which
    	    \begin{align}
    	    \vec{d}(\bm{k}) = t[\sin(k_x),\sin(k_y),m+\cos(k_x)+\cos(k_y)],
    	    \end{align}
    	    with hopping strength $t$ and mass term $m$. Then
    	    \begin{align}
    	    \epsilon_{\pm\pm}^u = \pm\Gamma\sqrt{\frac{(m\!+\!2)^2t^2}{\Gamma^2}-2\pm 2\sqrt{1-\frac{(m\!+\!2)^2t^2}{\Gamma^2}}}
    	    \end{align}
    	    at $\k = 0$, and so we identify $\Gamma_c=(m+2)t$
    	    at and above which isolated flat bands form. Fig.~\ref{fig:qwz} depicts the resulting band structure as a function of $\Gamma$ for $m=1$, for which the Chern number in the zero dissipation limit is 1. As the band width is maximal at $\k=0$, for $\Gamma < \Gamma_c$ the dispersion can be seen to vanish already in regions of the Brillouin zone (BZ) with narrower band gap.
    	    The lifetime, $\tau=\hbar/|\text{Im}(\epsilon)|$, decreases as $1/\Gamma$ for $\epsilon_{\pm-}^u$ and grows as $\Gamma$ for $\epsilon_{\pm+}^u$, which follows from Taylor expansion of the square root. Consequently, a stable flat band spanning the entire BZ emerges for strong dissipation $\Gamma \gg \Gamma_c$. We note that the Berry connection vanishes for $\Gamma > \Gamma_c$ and the Chern number is zero [see Supplemental Material]. A general classification of the topology of these flat bands remains an open question.
    	    

    \section{Example: Spinful One-Band Model with TRS}
    	    \begin{figure}
    	        \includegraphics[width=\columnwidth]{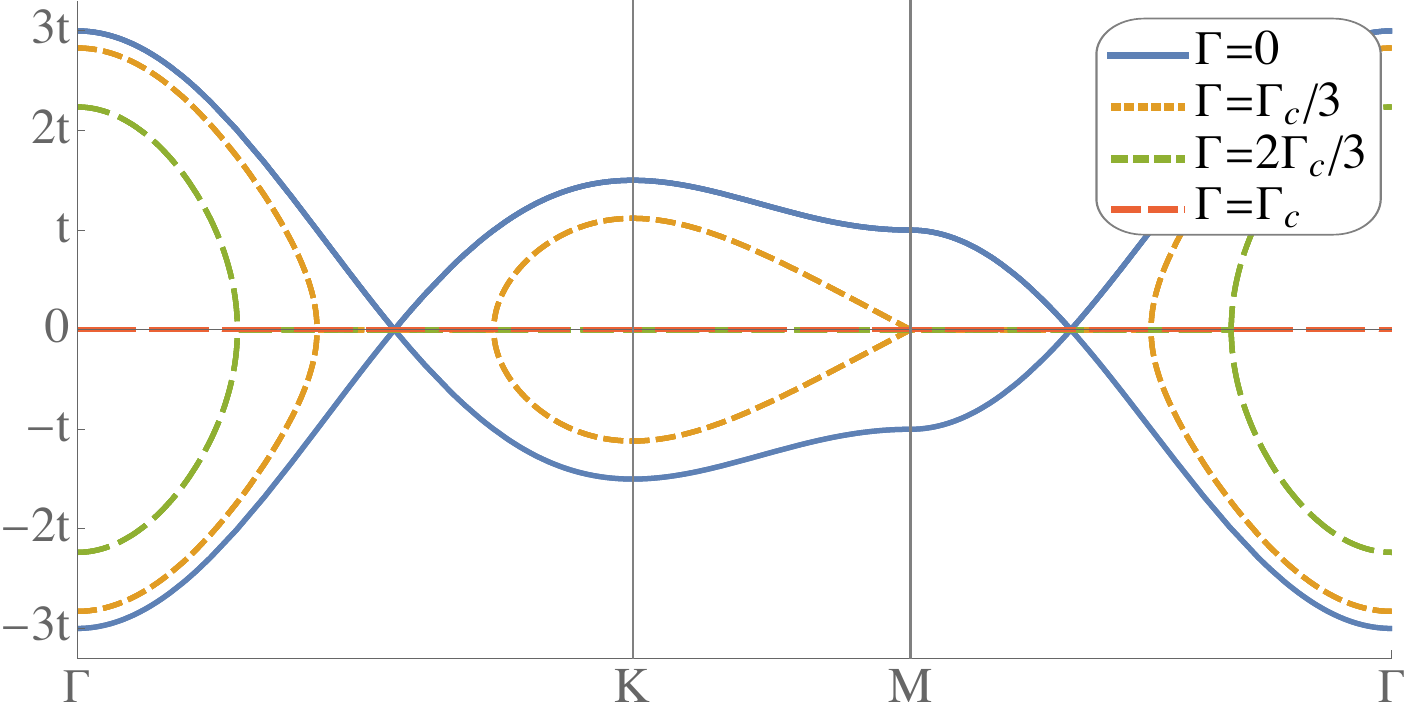}
    	        \caption{\textbf{Band flattening by dissipation in a spinful one-band model on a triangular lattice} with nearest neighbor hopping terms for a variety of dissipation rates, corresponding for instance to WSe$_2$ bilayers. 
    	        Both electron and hole bands are plotted.}
    	        \label{fig:1-band}
    	    \end{figure}
    	    Consider now the general time-reversal symmetric, inversion symmetric Hamiltonian for a single band with spin $H(\bm{k}) = d(\bm{k})\sigma_0$
    	    where $d(\bm{k})=d(-\bm{k})$. A straightforward spinful generalization of Eq. (\ref{eq:jumpSimple}) is of Majorana form
    	    \begin{align}\label{eq:non-trivial-spinful-dissipation}
    	    J_\sigma = c_{\bm{k},\sigma}+c_{-\bm{k},-\sigma}^\dagger,
    	    \end{align}
    	    with $\sigma=\ \uparrow$ \textit{and} $\downarrow$, and taking both $J_\uparrow$ and $J_\downarrow$ to ensure time-reversal symmetry. The modes of $L$ are the same as in Eq. (\ref{eq:d.sigma-modes}) but with $|\vec{d}|^2\mapsto |d|^2$.
    	    We see that the real part is zero for $\Gamma^2\geq |d(\bm{k})|^2$, which defines a critical
    	    $
    	    \Gamma_c = \max_{\bm{k},\pm} |d(\bm{k})|
    	    $,
    	    above which flat bands form. As in the spinless model, half of the modes are long lived and half are short lived. These modes remain stable when adding a U(1) conserving spin-orbit coupling term $\lambda(\bm{k})\sigma_3$.
    	    
    	    Notably, these flat bands emerge in the large $\Gamma$ limit which is most reasonably obtained when the electronic bandwidth is small compared to the system-bath coupling. This suggests TMD moir\'e heterostructures with a twist-tunable bandwidth as a platform for realization. For instance, a system with an isolated energy band near the Fermi energy with a bandwidth $\sim10$ meV is twisted bilayer WSe$_2$ with a twist angle $\sim 2^\circ$ \cite{wang2020correlated,devakul2021}.
    	    The moir\'e superlattice of this system is triangular, so the nearest-neighbor hopping tight binding model takes the form
    	    \begin{align}
    	    d(\bm{k}) &= t (\cos(k_1)+\cos(k_2)+\cos(k_1-k_2)),
    	    \end{align}
    	    where $\epsilon_0$, and $t$ set the energy scales of the model, $k_i=k\cdot a_i$ and $a_1=(1,0)a$, $a_2=(1/2,\sqrt{3}/2)a$ and lattice constant $a$.
    	    Fig. \ref{fig:1-band} shows the influence of the dissipation on the band structure of the quadratic Lindbladian. As expected, the band flattens with increasing dissipation strength and is dispersionless for all $\bm{k}$ above $\Gamma_c$.
        
        \section{Superconducting Substrates}
            Having revealed a generic mechanism for flat band formation in Lindbladian systems, we see that the essential property to engineer the substrate is that jump operators obey the form Eq. (\ref{eq:JumpOpLinear}) with both $a$ and $b$ non-zero. For electrons, a superconductor is such a substrate, and the simplest scenario to realize these flat bands entails a 2D material with an isolated band layered on top of a bulk superconductor.
            
            The resulting jump processes involve electrons/holes jumping from the system into the superconducting bulk. If the bandwidth is large compared to the superconducting gap, flat bands will emerge, however with shortened lifetimes for electronic states above the superconducting gap. If, however, the electronic band lies within the substrate's superconducting gap, jump processes from sufficiently strong system-substrate hybridization arise solely through particle exchange processes via intermediate pair breaking states. For states of the 2D material near the center of the superconducting gap, these processes retain approximately equal electron and hole amplitudes. The corresponding jump operator thus takes the form of Eq. (\ref{eq:non-trivial-spinful-dissipation}) and realizes a flat band with long lifetime.

            %
            
        
        \section{Conclusion}
            We have shown a new and robust route for realizing dispersionless bands in generic low-dimensional quantum systems via strong coupling to a bath. The underlying mechanism can be traced to the emergence of a dissipationless dark space of the system-bath coupling, and applies generically for systems with an odd number of valence bands with or without spin. We argued that in a fermionic system a superconductor could serve as such a bath. An intriguing follow-up question concerns whether analogous considerations can be utilized for bosons, such as photonic systems coupled to a squeezed bath \cite{walls1983squeezed,vaccaro1989phase}. More broadly, flat bands can host a panoply of interesting unconventional quantum phases, as electronic behavior becomes exclusively governed by Coulomb interactions. Our results suggest substrate engineering as a promising alternate route towards formation or control of these phases.
    
    \section{Acknowledgments}
        S.T. and M.C. acknowledge support from the NSF under Grant No. DGE-1845298, and Grant No. DMR-2132591 respectively.

%




    
    \supplement
    \renewcommand{\section}[1]{\begin{center}\large \textbf{#1}\end{center}}
    \widetext
    \setlength{\parindent}{0pt}
    
    \title{Supplemental Material: Dissipation Induced Flat Bands}
    
    \author{Spenser Talkington}
    \affiliation{Department of Physics and Astronomy, University of Pennsylvania, Philadelphia, Pennsylvania 19104, USA}
    \email{spenser@upenn.edu}
    
    \author{Martin Claassen}
    \affiliation{Department of Physics and Astronomy, University of Pennsylvania, Philadelphia, Pennsylvania 19104, USA}
    \email{claassen@upenn.edu}
    
    \maketitle
    
    
    \section{SI. Derivation of the Single-Particle Lindbladian}\label{section:}
        
        \subsection{Preliminaries}
    
        The Lindbladian is \cite{gardiner2004quantum}
        	\begin{align}
        	\mathcal{L}[\rho] := [H,\rho] - \frac{i\Gamma}{2} \sum_m \left(\{J_m^\dagger J_m,\rho\} - 2 J_m\rho J_m^\dagger\right),
        	\end{align}
        	where we take the coherent part of the evolution to be given by a Bloch Hamiltonian
        	\begin{align}
            H(\bm{k}) := \sum_{\alpha,\beta} h_{\alpha\beta}(\bm{k}) c_{\bm{k},\alpha}^\dagger c_{\bm{k},\beta},
        	\end{align}
        	and the jump operators are given by:
        	\begin{align}
        	J_m(\bm{k}) := \sum_\alpha a_{m,\alpha}(\bm{k})c_{\bm{k},\alpha} + b_{m,\alpha}(\bm{k})c_{\bm{k},\alpha}^\dagger,
        	\end{align}
        	both of which are expressed in terms of sum(s) over orbitals $\alpha$ ($\beta$).
        	\\
        	
        	This can be reexpressed in vectorized form \cite{amshallem2015three}
        	\begin{align}
        	\hat{\mathcal{L}} = (\1\otimes H - H^\top\otimes \1) -\frac{i\Gamma}{2}\sum_m\left(\1\otimes J_m^\dagger J_m + (J_m^\dagger J_m)^\top\otimes \1 - 2 J_m^*\otimes J_m\right),
        	\end{align}
        	or, using the hermiticity of $H$ and $J_m^\dagger J_m$,
        	\begin{align}
        	\hat{\mathcal{L}} = (\1\otimes H - H^*\otimes \1) -\frac{i\Gamma}{2}\sum_m\left(\1\otimes J_m^\dagger J_m + (J_m^\dagger J_m)^*\otimes \1 - 2 J_m^*\otimes J_m\right).
        	\end{align}
        	
        	Below, we show that
        	\begin{align}
        	\hat{\mathcal{L}} = (H_L-H_R) - \frac{i\Gamma}{2}\sum_m \left(J_{m,L}^\dagger J_{m,L} + (J_{m,R}^\dagger J_{m,R})^* - 2 J_{m,L}J_{m,R}^* \right),
        	\end{align}
        	where
        	\begin{align}
        	H_L &:= \sum_{\alpha,\beta} h_{\alpha\beta} \ell_\alpha^\dagger \ell_\beta,\\
        	H_R &:= \sum_{\alpha,\beta} h_{\alpha\beta} r_\beta^\dagger r_\alpha
        	= \sum_{\alpha,\beta} h_{\beta\alpha} r_\alpha^\dagger r_\beta,\\
        	J_{m,L} &:= \sum_\alpha a_{m,\alpha}\ell_\alpha + b_{m,\alpha}\ell_\alpha^\dagger,\\
        	J_{m,R} &:= \sum_{\alpha} - a_{m,\alpha}r_\alpha + b_{m,\alpha}r_\alpha^\dagger.
        	\end{align}
        	
        	The ``left superfermions" and ``right superfermions" are
        	\begin{align}
        	\ell_j := \mathcal{P}\otimes c_j,\\
            \ell_j^\dagger := \mathcal{P}\otimes c_j^\dagger,\\
            r_j:=c_j \mathcal{P}\otimes\1,\\
            r_j^\dagger:=\mathcal{P} c_j^\dagger\otimes\1,
        	\end{align}
        	Compared to the manuscript $r\leftrightarrow r^\dagger$ we first arrive at a ``block" form of the Lindbladian that makes its dark space zero modes manifest. In Section SI E we transform it into the BdG form presented in the manuscript.
        	
        	The parity operator is
        	\begin{align}
        	\mathcal{P} := e^{i\pi \mathcal{N}}
        	= (-\1)^{\mathcal{N}},
        	\end{align}
        	where the number operator is the non-local operator
        	\begin{align}
        	\mathcal{N} := \sum_j c_j^\dagger c_j,
        	\end{align}
        	and we have for all $j$
        	\begin{align}
        	    \{c_j,\mathcal{P}\}&=0,\\
        	    \{c_j^\dagger,\mathcal{P}\}&=0,
        	\end{align}
        	which motivates the definition of parity.
        	
        	We also have the standard Fermionic anti-commutation relations
        	\begin{align}
        	    \{l_j,r_{j'}\}&=0,\\
        	    \{l_j,r_{j'}^\dagger\}&=0,\\
        	    \{l_j^\dagger,r_{j'}\}&=0,\\
        	    \{l_j^\dagger,r_{j'}^\dagger\}&=0,\\
        	    \{l_j,l_{j'}\}&=\delta_{j,j'},\\
        	    \{r_j,r_{j'}^\dagger\}&=\delta_{j,j'}.
        	\end{align}

        \subsection{Coherent Part}
    
        	Now,
        	\begin{align}
        	H_L &= \sum_{\alpha,\beta} h_{\alpha\beta} \ell_\alpha^\dagger \ell_\beta,\\
        	&= \sum_{\alpha,\beta} h_{\alpha\beta} (\mathcal{P}\otimes c_\alpha^\dagger)(\mathcal{P}\otimes c_\beta),\\
        	&= \sum_{\alpha,\beta} h_{\alpha\beta}\, \1\otimes c_\alpha^\dagger c_\beta,\\
        	&= \1 \otimes \sum_{\alpha,\beta} h_{\alpha\beta} c_\alpha^\dagger c_\beta,\\
        	&= \1\otimes H,
        	\end{align}
        	and
        	\begin{align}
        	H_R &= \sum_{\alpha,\beta} h_{\alpha\beta} r_\beta^\dagger r_\alpha,\\
        	&= \sum_{\alpha,\beta} h_{\alpha\beta} (\mathcal{P}c_\beta^\dagger\otimes \1)(c_\alpha \mathcal{P}\otimes \1),\\
        	&= \sum_{\alpha,\beta} h_{\alpha\beta}\, \mathcal{P}c_\beta^\dagger c_\alpha \mathcal{P}\otimes \1^2,\\
        	&= \sum_{\alpha,\beta} h_{\alpha\beta}\, (- \1)^2\mathcal{P}^2c_\beta^\dagger c_\alpha \otimes \1^2,\\
        	&= \big(\sum_{\alpha,\beta} h_{\alpha\beta}\, c_\beta^\dagger c_\alpha\big) \otimes \1,\\
        	&= H^\top \otimes \1,\\
        	&= H^* \otimes \1.
        	\end{align}
        
        \subsection{Dissipative Part}
        
        	We have
        	\begin{align}
        	J_{m,L}^\dagger J_{m,L} &= \big(\sum_\alpha a_{m,\alpha}\ell_\alpha + b_{m,\alpha}\ell_\alpha^\dagger\big)^\dagger \big(\sum_\beta a_{m,\beta}\ell_\beta + b_{m,\beta}\ell_\beta^\dagger\big),\\
        	&= \sum_{\alpha,\beta} \big(a_{m,\alpha}^\dagger\ell_\alpha^\dagger + b_{m,\alpha}^\dagger\ell_\alpha\big) \big(a_{m,\beta}\ell_\beta + b_{m,\beta}\ell_\beta^\dagger\big),\\
        	&= \sum_{\alpha,\beta} a_{m,\alpha}^\dagger a_{m,\beta}\ell_\alpha^\dagger\ell_\beta + a_{m,\alpha}^\dagger b_{m,\beta}\ell_\alpha^\dagger\ell_\beta^\dagger + b_{m,\alpha}^\dagger a_{m,\beta}\ell_\alpha\ell_\beta + b_{m,\alpha}^\dagger b_{m,\beta} \ell_\alpha \ell_\beta^\dagger,\\
        	&= \sum_{\alpha,\beta} a_{m,\alpha}^\dagger a_{m,\beta}(\mathcal{P}\otimes c_\alpha^\dagger)(\mathcal{P}\otimes c_\beta) + a_{m,\alpha}^\dagger b_{m,\beta}(\mathcal{P}\otimes c_\alpha^\dagger)(\mathcal{P}\otimes c_\beta^\dagger), \\&\qquad + b_{m,\alpha}^\dagger a_{m,\beta}(\mathcal{P}\otimes c_\alpha)(\mathcal{P}\otimes c_\beta) + b_{m,\alpha}^\dagger b_{m,\beta} (\mathcal{P}\otimes c_\alpha) (\mathcal{P}\otimes c_\beta^\dagger),\\
        	&= \sum_{\alpha,\beta} a_{m,\alpha}^\dagger a_{m,\beta}(\1\otimes c_\alpha^\dagger c_\beta) + a_{m,\alpha}^\dagger b_{m,\beta}(\1\otimes c_\alpha^\dagger c_\beta^\dagger), \\&\qquad + b_{m,\alpha}^\dagger a_{m,\beta}(\1\otimes c_\alpha c_\beta) + b_{m,\alpha}^\dagger b_{m,\beta} (\1\otimes c_\alpha c_\beta^\dagger),\\
        	&= \1 \otimes \big (\sum_{\alpha,\beta} a_{m,\alpha}^\dagger a_{m,\beta} c_\alpha^\dagger c_\beta + a_{m,\alpha}^\dagger b_{m,\beta} c_\alpha^\dagger c_\beta^\dagger + b_{m,\alpha}^\dagger a_{m,\beta} c_\alpha c_\beta + b_{m,\alpha}^\dagger b_{m,\beta} c_\alpha c_\beta^\dagger \big),
        	\\
        	&= \1 \otimes \big (\sum_{\alpha} a_{m,\alpha}^\dagger c_\alpha^\dagger + b_{m,\alpha}^\dagger c_\alpha\big)\big(\sum_\beta a_{m,\beta} c_\beta + b_{m,\beta} c_\beta^\dagger \big),\\
        	&= \1 \otimes J_m^\dagger J_m,
        	\end{align}
        	and
        	\begin{align}
        	(J_{m,R}^\dagger J_{m,R})^* &= \big(\sum_{\alpha} - a_{m,\alpha}r_\alpha + b_{m,\alpha}r_\alpha^\dagger\big)^{\dagger,*} \big(\sum_{\beta} - a_{m,\beta}r_\beta + b_{m,\beta}r_\beta^\dagger\big)^*,\\
        	&= \big(\sum_{\alpha} - a_{m,\alpha}r_\alpha^{\dagger,*} + b_{m,\alpha}r_\alpha^*\big) \big(\sum_{\beta} - a_{m,\beta}^*r_\beta^* + b_{m,\beta}^*r_\beta^{\dagger,*}\big),\\
        	&= \sum_{\alpha,\beta} a_{m,\alpha}a_{m,\beta}^* r_\alpha^{\dagger,*} r_\beta^* - a_{m,\alpha} b_{m,\beta}^* r_{\alpha}^{\dagger,*} r_{\beta}^{\dagger,*} - b_{m,\alpha}a_{m,\beta}^*r_{\alpha}^* r_\beta^* + b_{m,\alpha}b_{m,\beta}^* r_\alpha^* r_\beta^{\dagger,*},\\
        	&= \sum_{\alpha,\beta} a_{m,\alpha}a_{m,\beta}^* (\mathcal{P}c_\alpha^\dagger\otimes\1)^* (c_\beta \mathcal{P}\otimes\1)^* - a_{m,\alpha} b_{m,\beta}^* (\mathcal{P}c_\alpha^\dagger\otimes\1)^* (\mathcal{P}c_\beta^\dagger\otimes\1)^*,\\&\qquad - b_{m,\alpha}a_{m,\beta}^*(c_\alpha \mathcal{P}\otimes\1)^* (c_\beta \mathcal{P}\otimes\1)^* + b_{m,\alpha}b_{m,\beta}^* (c_\alpha \mathcal{P}\otimes\1)^* (\mathcal{P}c_\beta^\dagger\otimes\1)^*,\\
        	&= \bigg(\sum_{\alpha,\beta} a_{m,\alpha}^*a_{m,\beta} (\mathcal{P}c_\alpha^\dagger\otimes\1) (c_\beta \mathcal{P}\otimes\1) - a_{m,\alpha}^* b_{m,\beta} (\mathcal{P}c_\alpha^\dagger\otimes\1) (\mathcal{P}c_\beta^\dagger\otimes\1),\\&\qquad - b_{m,\alpha}^*a_{m,\beta}(c_\alpha \mathcal{P}\otimes\1) (c_\beta \mathcal{P}\otimes\1) + b_{m,\alpha}^*b_{m,\beta} (c_\alpha \mathcal{P}\otimes\1) (\mathcal{P}c_\beta^\dagger\otimes\1)\bigg)^*,\\
        	&= \bigg(\sum_{\alpha,\beta} a_{m,\alpha}^*a_{m,\beta} \mathcal{P}c_\alpha^\dagger c_\beta \mathcal{P} - a_{m,\alpha}^* b_{m,\beta} \mathcal{P}c_\alpha^\dagger \mathcal{P}c_\beta^\dagger - b_{m,\alpha}^*a_{m,\beta}c_\alpha \mathcal{P} c_\beta \mathcal{P} + b_{m,\alpha}^*b_{m,\beta} c_\alpha \mathcal{P}\mathcal{P}c_\beta^\dagger\bigg)^*\otimes \1,\\
        	&= \bigg(\sum_{\alpha,\beta} a_{m,\alpha}^*a_{m,\beta} c_\alpha^\dagger c_\beta + a_{m,\alpha}^* b_{m,\beta} c_\alpha^\dagger c_\beta^\dagger + b_{m,\alpha}^*a_{m,\beta}c_\alpha c_\beta + b_{m,\alpha}^*b_{m,\beta} c_\alpha c_\beta^\dagger\bigg)^*\otimes \1,\\
        	&= \bigg(\big(\sum_{\alpha}a_{m,\alpha}^*c_\alpha^\dagger + b_{m,\alpha}^*c_\alpha \big)\big(\sum_\beta a_{m,\beta}c_\beta + b_{m,\beta}c_\beta^\dagger\big)\bigg)^*\otimes \1,\\
        	&= (J_m^\dagger J_m)^*\otimes \1,
        	\end{align}
        	and
        	\begin{align}
        	J_{m,L}J_{m,R}^* &= \big(\sum_\alpha a_{m,\alpha}\ell_\alpha + b_{m,\alpha}\ell_\alpha^\dagger\big)\big(\sum_{\beta} - a_{m,\beta}r_\beta + b_{m,\beta}r_\beta^{\dagger}\big)^*,\\
        	&= \big(\sum_\alpha a_{m,\alpha}\ell_\alpha + b_{m,\alpha}\ell_\alpha^\dagger\big)\big(\sum_{\beta} - a_{m,\beta}^*r_\beta^* + b_{m,\beta}^*r_\beta^{\dagger,*}\big),\\
        	&= \sum_{\alpha,\beta} - a_{m,\alpha}a_{m,\beta}^*\ell_\alpha r_\beta^* + a_{m,\alpha}b_{m,\beta}^*\ell_\alpha r_\beta^{\dagger,*} - b_{m,\alpha}a_{m,\beta}^*\ell_\alpha^\dagger r_\beta^* + b_{m,\alpha}b_{m,\beta}^*\ell_\alpha^\dagger r_\beta^{\dagger,*},\\
        	&= \sum_{\alpha,\beta} - a_{m,\alpha}a_{m,\beta}^*(\mathcal{P}\otimes c_\alpha) (c_\beta \mathcal{P}\otimes \1)^* + a_{m,\alpha}b_{m,\beta}^*(\mathcal{P}\otimes c_\alpha) (\mathcal{P}c_\beta^\dagger\otimes\1)^*,\\&\qquad - b_{m,\alpha}a_{m,\beta}^*(\mathcal{P}\otimes c_\alpha^\dagger) (c_\beta \mathcal{P}\otimes \1)^* + b_{m,\alpha}b_{m,\beta}^*(\mathcal{P}\otimes c_\alpha^\dagger) (\mathcal{P}c_\beta^\dagger\otimes \1)^*,\\
        	&= \sum_{\alpha,\beta} - a_{m,\alpha}a_{m,\beta}^*(\mathcal{P}(c_\beta \mathcal{P})^*\otimes c_\alpha) + a_{m,\alpha}b_{m,\beta}^*(\mathcal{P}(\mathcal{P}c_\beta^\dagger)^*\otimes c_\alpha), \\&\qquad - b_{m,\alpha}a_{m,\beta}^*(\mathcal{P}(c_\beta \mathcal{P})^*\otimes c_\alpha^\dagger) + b_{m,\alpha}b_{m,\beta}^*(\mathcal{P}(\mathcal{P}c_\beta^\dagger)^*\otimes c_\alpha^\dagger),\\
        	&= \sum_{\alpha,\beta} a_{m,\alpha}a_{m,\beta}^*((c_\beta)^*\otimes c_\alpha) + a_{m,\alpha}b_{m,\beta}^*((c_\beta^\dagger)^*\otimes c_\alpha), \\&\qquad + b_{m,\alpha}a_{m,\beta}^*((c_\beta)^*\otimes c_\alpha^\dagger) + b_{m,\alpha}b_{m,\beta}^*((c_\beta^\dagger)^*\otimes c_\alpha^\dagger),\\
        	&= \big(\sum_\beta a_{m,\beta}^* (c_\beta)^* + b_{m,\beta}^* (c_\beta^\dagger)^*\big)\otimes \big(\sum_\alpha a_{m,\alpha} c_\alpha + b_{m,\alpha} c_\alpha^\dagger\big),\\
        	&= \big(\sum_\beta a_{m,\beta} c_\beta + b_{m,\beta}c_\beta^\dagger\big)^*\otimes \big(\sum_\alpha a_{m,\alpha} c_\alpha + b_{m,\alpha} c_\alpha^\dagger\big),\\
        	&= J_m^* \otimes J_m.
        	\end{align}
    
    \subsection{Matrix Representation (Block Form)}
    
        In the last section, we showed that
    	\begin{align}
    	\hat{\mathcal{L}} = (H_L-H_R) - \frac{i\Gamma}{2}\sum_m \left(J_{m,L}^\dagger J_{m,L} + (J_{m,R}^\dagger J_{m,R})^* - 2 J_{m,L}J_{m,R}^* \right),
    	\end{align}
    	where
    	\begin{align}
    	H_L &:= \sum_{\alpha,\beta} h_{\alpha\beta} \ell_\alpha^\dagger \ell_\beta,\\
    	H_R &:= \sum_{\alpha,\beta} h_{\alpha\beta} r_\beta^\dagger r_\alpha
    	= \sum_{\alpha,\beta} h_{\beta\alpha} r_\alpha^\dagger r_\beta,\\
    	J_{m,L} &:= \sum_\alpha a_{m,\alpha}\ell_\alpha + b_{m,\alpha}\ell_\alpha^\dagger,\\
    	J_{m,R} &:= \sum_{\alpha} - a_{m,\alpha}r_\alpha + b_{m,\alpha}r_\alpha^\dagger.
    	\end{align}
    	With the Fermionic anti-commutation relations this is
    	\begin{align}
    	\hat{\mathcal{L}} = (H_L-H_R) - \frac{i\Gamma}{2}\sum_m \left(J_{m,L}^\dagger J_{m,L} + (J_{m,R}^\dagger J_{m,R})^* - J_{m,L}J_{m,R}^* + J_{m,R}^*J_{m,L} \right).
    	\end{align}
    	
    	We have
    	\begin{align}
    	\hat{\mathcal{L}} = \hat{\mathcal{L}}_\text{coh} + \hat{\mathcal{L}}_\text{dis}.
    	\end{align}
    	
    	And we introduce the notation for vectors of left and right creation and annihilation operators
    	\begin{align}
    	\bm{\ell} = (\ell_1, \ell_2, \dots, \ell_\text{num orbitals}),\\
    	\bm{\ell}^\dagger = (\ell_1^\dagger, \ell_2^\dagger, \dots, \ell_\text{num orbitals}^\dagger),\\
    	\bm{r} = (r_1, r_2, \dots, r_\text{num orbitals}),\\
    	\bm{r}^\dagger = (r_1^\dagger, r_2^\dagger, \dots, r_\text{num orbitals}^\dagger).
    	\end{align}
    	
    	Now the coherent part of the Lindbladian is
    	\begin{align}
    	\hat{\mathcal{L}}_\text{coh} &= (H_L - H_R),\\
    	&= \frac{1}{2}(H_L + H_L^\dagger - H_R - H_R^\dagger),\\
    	&= \frac{1}{2}(\bm{\ell}^\dagger H \bm{\ell} - \bm{\ell} H^\top \bm{\ell}^\dagger - \bm{r}^\dagger H^\top \bm{r} + \bm{r} H \bm{r}^\dagger),\\
    	&= \frac{1}{2} \begin{pmatrix}
    	\bm{\ell}^\dagger\\-\bm{r}\\\bm{r}^\dagger\\\bm{\ell}
    	\end{pmatrix}
    	\begin{pmatrix}
    	H & 0 & 0 & 0\\
    	0 & H & 0 & 0\\
    	0 & 0 & -H^\top & 0\\
    	0 & 0 & 0 & -H^\top
    	\end{pmatrix}
    	\begin{pmatrix}
    	\bm{\ell}\\-\bm{r}^\dagger\\\bm{r}\\\bm{\ell}^\dagger
    	\end{pmatrix},
    	\end{align}
    	where we used the anti-commutation relations to make the indices appear in the correct order.
    	\\
    	
    	Next we introduce a notation for matrices $A,B,C$ as
    	\begin{align}
    	A = [A_{\alpha,\beta}]
    	,\qquad
    	B = [B_{\alpha,\beta}]
    	,\qquad
    	C = [C_{\alpha,\beta}],
    	\end{align}
    	where
    	\begin{align}
    	A_{\alpha,\beta} = \sum_m a_{m,\alpha}^*a_{m,\beta}
    	,\qquad
    	B_{\alpha,\beta} = \sum_m b_{m,\alpha}b_{m,\beta}^*
    	,\qquad
    	C_{\alpha,\beta} = \sum_m a_{m,\alpha}^*b_{m,\beta}.
    	\end{align}
    
    \newpage
        
        The dissipative part of the Lindbladian is
        \begin{align}
    	\hat{\mathcal{L}}_\text{dis} = - \frac{i\Gamma}{2}\sum_m &\left(J_{m,L}^\dagger J_{m,L} + (J_{m,R}^\dagger J_{m,R})^* - 2 J_{m,L}J_{m,R}^* \right),\\
    	= - \frac{i\Gamma}{2}\sum_m &\left(J_{m,L}^\dagger J_{m,L} + (J_{m,R}^\dagger J_{m,R})^* - J_{m,L}J_{m,R}^* + J_{m,R}^*J_{m,L} \right),\\
    	= - \frac{i\Gamma}{2}\sum_{n,\alpha,\beta} \bigg(&(a_{m,\alpha}^*\ell_\alpha^\dagger+b_{m,\alpha}^*\ell_\alpha)(a_{m,\beta}\ell_\beta+b_{m,\beta}\ell_\beta^\dagger)\\ + \big(&(- a_{m,\alpha}^*r_\alpha^\dagger+b_{m,\alpha}^*r_\alpha)(- a_{m,\beta}r_\beta+b_{m,\beta}r_\beta^\dagger)\big)^*\\ - &(a_{m,\alpha}\ell_\alpha + b_{m,\alpha}\ell_\alpha^\dagger)(- a_{m,\beta}r_\beta+b_{m,\beta}r_\beta^\dagger)^*\\
    	+ &(- a_{m,\alpha}r_\alpha+b_{m,\alpha}r_\alpha^\dagger)^*(a_{m,\beta}\ell_\beta + b_{m,\beta}\ell_\beta^\dagger) \bigg),\\
    	= - \frac{i\Gamma}{2}\sum_{n,\alpha,\beta} \bigg(&(a_{m,\alpha}^*a_{m,\beta}\ell_\alpha^\dagger\ell_\beta+a_{m,\alpha}^*b_{m,\beta}\ell_\alpha^\dagger\ell_\beta^\dagger+b_{m,\alpha}^*a_{m,\beta}\ell_\alpha\ell_\beta+b_{m,\alpha}^*b_{m,\beta}\ell_\alpha\ell_\beta^\dagger)\\ + &(a_{m,\alpha}a_{m,\beta}^* r_\alpha^\dagger r_\beta - a_{m,\alpha}b_{m,\beta}^* r_\alpha^\dagger r_\beta^\dagger- b_{m,\alpha}a_{m,\beta}^*r_\alpha r_\beta +b_{m,\alpha} b_{m,\beta}^*r_\alpha r_\beta^\dagger)\\ - &(- a_{m,\alpha}a_{m,\beta}^*\ell_\alpha r_\beta + a_{m,\alpha} b_{m,\beta}^* \ell_\alpha r_\beta^\dagger - b_{m,\alpha}a_{m,\beta}^* \ell_\alpha^\dagger r_\beta+b_{m,\alpha} b_{m,\beta}^* \ell_\alpha^\dagger r_\beta^\dagger)\\
    	+ &(- a_{m,\alpha}^*a_{m,\beta}r_\alpha \ell_\beta - a_{m,\alpha}^* b_{m,\beta} r_\alpha \ell_\beta^\dagger + b_{m,\alpha}^*a_{m,\beta} r_\alpha^\dagger\ell_\beta+b_{m,\alpha}^* b_{m,\beta} r_\alpha^\dagger \ell_\beta^\dagger) \bigg),\\
    	= - \frac{i\Gamma}{2}\big(&(\bm{\ell}^\dagger A\bm{\ell}+\bm{\ell}^\dagger C\bm{\ell}^\dagger+\bm{\ell}C^\dagger\bm{\ell}+\bm{\ell}B^*\bm{\ell}^\dagger)\\ + &(\bm{r}^\dagger A^*\bm{r} - \bm{r}^\dagger C^* \bm{r}^\dagger - \bm{r}C^\top\bm{r} + \bm{r}B \bm{r}^\dagger)\\ - &(- \bm{\ell}A^* \bm{r} + \bm{\ell}C^* \bm{r}^\dagger - \bm{\ell}^\dagger C^\top \bm{r} + \bm{\ell}^\dagger B \bm{r}^\dagger)\\
    	+ &(- \bm{r}A \bm{\ell} - \bm{r}C \bm{\ell}^\dagger + \bm{r}^\dagger C^\dagger \bm{\ell} + \bm{r}^\dagger B^* \bm{\ell}^\dagger) \big),\\
    	= -\frac{i\Gamma}{2} \big(&\bm{\ell}^\dagger A \bm{\ell} - \bm{\ell}^\dagger B \bm{r}^\dagger + \bm{\ell}^\dagger C^\top \bm{r} + \bm{\ell}^\dagger C \bm{\ell}^\dagger\\
    	- &\bm{r} A \bm{\ell} + \bm{r} B \bm{r}^\dagger - \bm{r} C^\top \bm{r} - \bm{r} C \bm{\ell}^\dagger\\
    	+ &\bm{r}^\dagger C^\dagger \bm{\ell} - \bm{r}^\dagger C^* \bm{r}^\dagger + \bm{r}^\dagger A^* \bm{r} + \bm{r}^\dagger B^* \bm{\ell}^\dagger\\
    	+ &\bm{\ell} C^\dagger \bm{\ell} - \bm{\ell} C^* \bm{r}^\dagger + \bm{\ell} A^* \bm{r} + \bm{\ell} B^* \bm{\ell}^\dagger\big),\\
    	= - \frac{i\Gamma}{2}
    	\begin{pmatrix}
    	\bm{\ell}^\dagger\\-\bm{r}\\\bm{r}^\dagger\\\bm{\ell}
    	\end{pmatrix}
    	&\begin{pmatrix}
    	A & B & C^\top & C\\
    	A & B & C^\top & C\\
    	C^\dagger & C^* & A^* & B^*\\
    	C^\dagger & C^* & A^* & B^*
    	\end{pmatrix}
    	\begin{pmatrix}
    	\bm{\ell}\\-\bm{r}^\dagger\\\bm{r}\\\bm{\ell}^\dagger
    	\end{pmatrix},\\
    	= - \frac{i\Gamma}{2}
    	&\begin{pmatrix}
    	\bm{\ell}^\dagger\\-\bm{r}\\\bm{r}^\dagger\\\bm{\ell}
    	\end{pmatrix}
    	\begin{pmatrix}
    	A & B & C^\top & C\\
    	A & B & C^\top & C\\
    	C^\dagger & C^* & A^* & B^*\\
    	C^\dagger & C^* & A^* & B^*
    	\end{pmatrix}
    	\begin{pmatrix}
    	\bm{\ell}\\-\bm{r}^\dagger\\\bm{r}\\\bm{\ell}^\dagger
    	\end{pmatrix}.
    	\end{align}
    	Now, it turns out that the normal modes in this matrix representation are \textit{not} normal modes of the second quantized Lindbladian $\mathcal{L}$. The matrix representation whose normal modes \textit{are} normal modes of the second quantized Lindbladian have $r\leftrightarrow r^\dagger$.
    	
    \newpage
    \subsection{Matrix Representation (BdG Form)}\label{sec:bdg-form}
    	
    	From the last subsection, we have that the vectorized Lindbladian permits a representation in terms of left and right fermions. Now, let us reexpress this in the form whose normal modes are the normal modes of the second quantized Lindbladian.
    	To do so we take $r\leftrightarrow r^\dagger$ so that
    	\begin{align}
    	    r_j &= \mathcal{P}c_j^\dagger\otimes\1,\\
    	    r_j^\dagger &= c_j\mathcal{P}\otimes\1.
    	\end{align}
    	Doing so, our ordered basis is now $(\bm{\ell},-\bm{r},\bm{r}^\dagger,\bm{\ell}^\dagger)$. Now, 
    	$U(\bm{\ell},\bm{r},\bm{\ell}^\dagger,+\bm{r}^\dagger)=(\bm{\ell},-\bm{r},+\bm{r}^\dagger,+\bm{\ell}^\dagger)$ for \begin{align}
    	U = \begin{pmatrix}
    	1&0&0&0\\0&-1&0&0\\0&0&0&1\\0&0&1&0
    	\end{pmatrix},
    	\end{align}
    	under which $L_\text{coh}$ is clearly invariant.
    	
    	Now,
    	\begin{align}
    	U^{-1}L_\text{dis}U = \begin{pmatrix}
    	A&-B&C&C^\top\\
    	-A&B&-C&-C^\top\\
    	C^\dagger & -C^* & B^* & A^*\\
    	C^\dagger & -C^* & B^* & A^*
    	\end{pmatrix}.
    	\end{align}
    	
    	This means that
    	\begin{align} \hat{\mathcal{L}}_\text{dis} &= -\frac{i\Gamma}{4}
    	\begin{pmatrix}
    	\bm{\ell}^\dagger\\\bm{r}^\dagger\\\bm{\ell}\\\bm{r}
    	\end{pmatrix}
    	\begin{pmatrix}
    	2A&-2B&2C&2C^\top\\
    	-2A&2B&-2C&-2C^\top\\
    	2C^\dagger & - 2C^* & 2B^* & 2A^*\\
    	2C^\dagger & -2C^* & 2B^* & 2A^*
    	\end{pmatrix}
    	\begin{pmatrix}
    	\bm{\ell}\\\bm{r}\\\bm{\ell}^\dagger\\\bm{r}^\dagger
    	\end{pmatrix},
    	\end{align}
    	or, using the anti-commutation relations
    	\begin{align}
    	2A\bm{\ell}^\dagger\bm{\ell} + 2B^*\bm{\ell}\bm{\ell}^\dagger &= (A- B^{*,\top})\bm{\ell}^\dagger\bm{\ell} + (B^*- A^\top)\bm{\ell}\bm{\ell}^\dagger,\\
    	2B\bm{r}^\dagger\bm{r} + 2A^* (+\bm{r})(+\bm{r}^\dagger) &= (B- A^{*,\top})\bm{r}^\dagger\bm{r} + (A^*- B^{\top}) (+\bm{r})(+\bm{r}^\dagger),\\
    	2C\bm{\ell}^\dagger\bm{\ell}^\dagger &= (C- C^\top)\bm{\ell}^\dagger\bm{\ell}^\dagger,\\
    	-2C^\top\bm{r}^\dagger(+ \bm{r}^\dagger) &= -(C^\top- C)\bm{r}^\dagger(+ \bm{r}^\dagger),\\
    	2C^\dagger \bm{\ell}\bm{\ell} &= (C^\dagger- C^*)\bm{\ell}\bm{\ell},\\
    	-2C^*(+\bm{r})\bm{r}&=(-C^*+ C^\dagger)(+\bm{r})\bm{r}.
    	\end{align}
    	We can symmetrize (using the Hermiticity of $A$ and $B$)
    	\begin{align} \hat{\mathcal{L}}_\text{dis} &= -\frac{i\Gamma}{4}
    	\begin{pmatrix}
    	\bm{\ell}^\dagger\\\bm{r}^\dagger\\\bm{\ell}\\\bm{r}
    	\end{pmatrix}
    	\begin{pmatrix}
    	A- B&-2B&C- C^\top&2C^\top\\
    	-2A&B- A&-2C&+ C-C^\top\\
    	C^\dagger- C^* & - 2C^* & B^*- A^\top & 2A^*\\
    	2C^\dagger & + C^\dagger-C^* & 2B^* & A^*- B^\top
    	\end{pmatrix}
    	\begin{pmatrix}
    	\bm{\ell}\\\bm{r}\\\bm{\ell}^\dagger\\\bm{r}^\dagger
    	\end{pmatrix},
    	\end{align}
    	or, simplifying
    	\begin{align}
        \hat{\mathcal{L}}_\text{dis} &= -\frac{i\Gamma}{4}
    	\begin{pmatrix}
    	\bm{\ell}^\dagger\\\bm{r}^\dagger\\\bm{\ell}\\\bm{r}
    	\end{pmatrix}
    	\begin{pmatrix}
    	A-B&-2B&C-C^\top&2C^\top\\
    	-2A&-(A-B)&-2C&C-C^\top\\
    	-(C-C^\top)^*&-2C^*&-(A-B)^\top & 2A^\top\\
    	2C^\dagger&-(C-C^\top)^*&2B^\top&(A-B)^\top
    	\end{pmatrix}
    	\begin{pmatrix}
    	\bm{\ell}\\\bm{r}\\\bm{\ell}^\dagger\\\bm{r}^\dagger
    	\end{pmatrix},
    	\end{align}
    	which, as desired is in BdG form.

    \newpage
    \section{SII. Pseudospin Representation of the Lindbladian}\label{appendix:pseudospin}
    
        Now introduce the pseudospins $\eta$ and $\tau$, where $\eta$ rotates particles and holes, and $\tau$ rotates left and right contours.
    	
    	In this representation, the coherent part is
    	\begin{align}
    	L_\text{coh} 
    	&= \frac{1}{2}[(\text{Re}(H_{\bm{k}}+H_{-\bm{k}}))
    	+ i(\text{Im}(H_{\bm{k}}-H_{-\bm{k}}))] \eta_3\otimes\tau_0\\
    	&+ \frac{1}{2}[(\text{Re}(H_{\bm{k}}-H_{-\bm{k}}))
    	+ i(\text{Im}(H_{\bm{k}}+H_{-\bm{k}}))]\eta_0\otimes\tau_0.
    	\end{align}
    	When $H:=H_{\bm{k}}=H_{-\bm{k}}$ this becomes
    	\begin{align}
    	L_\text{coh} = \Re(H)\eta_3\otimes\tau_0+i\,\Im(H)\eta_0\otimes\tau_0,
    	\end{align}
    	and the dissipative part is
    	\begin{align}\begin{split}
    	L_\text{dis} = \frac{\Gamma}{2}\frac{1}{2}\bigg[[\text{Re}(A_{\bm{k}}+A_{\bm{-k}}-B_{\bm{k}}-B_{\bm{-k}})+i\text{Im}(A_{\bm{k}}-A_{\bm{-k}}-B_{\bm{k}}+B_{\bm{-k}})]\eta_3\otimes\tau_3\\
    	+[\text{Re}(A_{\bm{k}}-A_{\bm{-k}}-B_{\bm{k}}+B_{\bm{-k}})+i\text{Im}(A_{\bm{k}}+A_{\bm{-k}}-B_{\bm{k}}-B_{\bm{-k}})]\eta_0\otimes\tau_3\\
    	-[\text{Re}(-A_{\bm{k}}+A_{-\bm{k}}+B_{\bm{k}}-B_{-\bm{k}})+i\text{Im}(-A_{\bm{k}}-A_{-\bm{k}}+B_{\bm{k}}+B_{-\bm{k}})]i\eta_3\otimes\tau_2\\
    	-[\text{Re}(-A_{\bm{k}}-A_{-\bm{k}}+B_{\bm{k}}+B_{-\bm{k}})+i\text{Im}(-A_{\bm{k}}+A_{-\bm{k}}+B_{\bm{k}}-B_{-\bm{k}})]i\eta_0\otimes\tau_2\\
    	-[\text{Re}(A_{\bm{k}}+A_{-\bm{k}}+B_{\bm{k}}+B_{-\bm{k}})+i\text{Im}(A_{\bm{k}}-A_{-\bm{k}}+B_{\bm{k}}-B_{-\bm{k}})]\eta_3\otimes\tau_1\\
    	-[\text{Re}(A_{\bm{k}}-A_{-\bm{k}}+B_{\bm{k}}-B_{-\bm{k}})+i\text{Im}(A_{\bm{k}}+A_{-\bm{k}}+B_{\bm{k}}+B_{-\bm{k}})]\eta_0\otimes\tau_1\\
    	+[\text{Re}(C_{\bm{k}}-C_{-\bm{k}}+C_{-\bm{k}}^\top-C_{\bm{k}}^\top)+i\text{Im}(C_{\bm{k}}+C_{-\bm{k}}+C_{-\bm{k}}^\top+C_{\bm{k}}^\top)]i\eta_1\otimes\tau_2\\
    	+[\text{Re}(-C_{\bm{k}}-C_{-\bm{k}}-C_{-\bm{k}}^\top-C_{\bm{k}}^\top)+i\text{Im}(-C_{\bm{k}}+C_{-\bm{k}}-C_{-\bm{k}}^\top+C_{\bm{k}}^\top)]\eta_2\otimes\tau_2\\
    	+[\text{Re}(-C_{\bm{k}}+C_{-\bm{k}}+C_{-\bm{k}}^\top-C_{\bm{k}}^\top)+i\text{Im}(-C_{\bm{k}}-C_{-\bm{k}}+C_{-\bm{k}}^\top+C_{\bm{k}}^\top)]i\eta_2\otimes\tau_1\\
    	+[\text{Re}(-C_{\bm{k}}-C_{-\bm{k}}+C_{-\bm{k}}^\top+C_{\bm{k}}^\top)+i\text{Im}(-C_{\bm{k}}+C_{-\bm{k}}+C_{-\bm{k}}^\top-C_{\bm{k}}^\top)]\eta_1\otimes\tau_1\\
        +[\text{Re}(C_{\bm{k}}-C_{-\bm{k}}-C_{-\bm{k}}^\top+C_{\bm{k}}^\top)+i\text{Im}(C_{\bm{k}}+C_{-\bm{k}}-C_{-\bm{k}}^\top-C_{\bm{k}}^\top)]\eta_1\otimes\tau_0\\
    	+[\text{Re}(C_{\bm{k}}+C_{-\bm{k}}-C_{-\bm{k}}^\top-C_{\bm{k}}^\top)+i\text{Im}(C_{\bm{k}}-C_{-\bm{k}}-C_{-\bm{k}}^\top+C_{\bm{k}}^\top)]i\eta_2\otimes\tau_0
    	\bigg].
    	\end{split}\end{align}
    	When $A:=A_{\bm{k}}=A_{-\bm{k}}$, $B:=B_{\bm{k}}=B_{-\bm{k}}$, $C:=C_{\bm{k}}=C_{-\bm{k}}$, $C^\top:=C^\top_{\bm{k}}=C^\top_{-\bm{k}}$, this becomes
    	\begin{align}\begin{split}
    	L_\text{dis} = \frac{\Gamma}{2}\big[&\Re\left(A-B\right)\eta_3\otimes\tau_3+i\, \Im\left(A-B\right)\eta_0\otimes\tau_3\\
    	-&\Im\left(A-B\right)\eta_3\otimes\tau_2 +i\,\Re\left(A-B\right)\eta_0\otimes\tau_2\\
    	- &\Re\left(A+B\right)\eta_3\otimes\tau_1 -i\,\Im\left(A+B\right)\eta_0\otimes\tau_1\\
    	-&\Im\left(C+C^\top\right)\eta_1\otimes\tau_2-\Re\left(C+C^\top\right)\eta_2\otimes\tau_2\\
    	-&\Re\left(C-C^\top\right)\eta_1\otimes\tau_1+\Im\left(C-C^\top\right)\eta_2\otimes\tau_1\\
    	+i\,&\Im\left(C-C^\top\right)\eta_1\otimes\tau_0 + i\,\Re\left(C-C^\top\right)\eta_2\otimes\tau_0\big].
    	\end{split}\end{align}
    	
    	Now with Hermitian symmetry operator
    	$
    	\mathcal{D} = \eta_3\otimes\tau_1
    	$
    	if
    	$
    	[\mathcal{D},\mathcal{L}_\text{dis}] = 0
    	$
    	then
    	\begin{align}
    	L_\text{dis} = \frac{\Gamma}{2}\bigg(-\Re\left(A+B\right)\eta_3\otimes\tau_1 -i\,\Im\left(A+B\right)\eta_0\otimes\tau_1
    	-\Im\left(C+C^\top\right)\eta_1\otimes\tau_2-\Re\left(C+C^\top\right)\eta_2\otimes\tau_2\bigg),
    	\end{align}
    	where we evaluated the commutators of the tensor products of the pseudospins.
    	Noting that with Hermitian symmetry $A=B$ and $C=C^\top$ this is
    	\begin{align}
    	L_\text{dis} = -\Gamma\bigg(\Re\left(A\right)\eta_3\otimes\tau_1 +i\,\Im\left(A\right)\eta_0\otimes\tau_1
    	+\Im\left(C\right)\eta_1\otimes\tau_2+\Re\left(C\right)\eta_2\otimes\tau_2\bigg).
    	\end{align}

    \newpage
    \section{SIII. Chern Number of the Dissipative QWZ Model}\label{appendix:chern}
        
        \subsection{Berry-Phase Topology}
            
            The Berry connection is
            \begin{align}
            \mathcal{A}_n^\mu = i\langle u_n^l|\partial_{k_\mu}|u_n^r\rangle.
            \end{align}
            where we use the \textit{left} and \textit{right} eigenvectors since $L$ in non-Hermitian and the left and right eigenvectors are not the same in general.\\
            
            The Berry curvature is
            \begin{align}
            \Omega_n^{\mu\nu} = \partial_{k_\mu}\mathcal{A}_n^\nu - \partial_{k_\nu}\mathcal{A}_n^\mu,
            \end{align}
            from which the first Chern number is
            \begin{align}
            c_n^{\mu\nu} = \frac{1}{2\pi} \int_{BZ}d\bm{k}\ \Omega_n^{\mu\nu}.
            \end{align}
            We will consider the Chern number $c_n^{xy}$ which is related to the Hall conductivity.
            \\
        
        \subsection{Model}
    	    
    	    We consider a Hamiltonian of the form
    	    \begin{align}
    	    H(\bm{k}) &= \vec{d}(\bm{k})\cdot\vec{\sigma},\\
    	    &= (d_1(\bm{k}),d_2(\bm{k}),d_3(\bm{k}))\cdot(\sigma_1,\sigma_2,\sigma_3).
    	    \end{align}
    	    
    	    Here we assume dissipation of the form
    	    \begin{align}
    	    J = c_{\bm{k},i}+c_{\bm{k},i}^\dagger,
    	    \end{align}
    	    for $i=1$ or $2$, and let a parameter $\Gamma$ set the rate of tunneling.
    	    
    	    As shown before, there are finite-lived energy bands at
    	    \begin{align}
    	    \epsilon_{\pm\pm}^u = \pm \frac{\Gamma}{2}\sqrt{|\vec{d}|^2/\Gamma^2-2 \pm 2\sqrt{1-|\vec{d}|^2/\Gamma^2}},
    	    \end{align}
    	    which are unique (as complex numbers). This non-degeneracy in complex-space suggests that a Berry-phase topological classification may be possible. Here, $|\vec{d}|=(d_1^2+d_2^2+d_3^2)^{1/2}$, and there are flat bands ($\Re(\epsilon^u_{\pm\pm})=0$) for
    	    \begin{align}
    	    \Gamma > \Gamma_c = |\vec{d}|,
    	    \end{align}
    	    which is the bandwidth of both bands in the dissipationless system.
    	    
    	    Now, we consider the Qi-Wu-Zhang model \cite{qi2006topological} in particular which has
    	    \begin{align}
    	    \vec{d} = t(\sin(k_x),\sin(k_y),m+\cos(k_x)+\cos(k_y)).
    	    \end{align}
    	    Proceeding numerically, we find that the Berry connection and curvature are non-zero for both the dissipationless and the dissipative bands, and as expected the band that is not dissipatively coupled to the bath coupled to the bath has the same Chern number as in the dissipationless case. We find that the Chern number of all bands with finite lifetimes is zero (even for small but non-zero dissipation rates), indicating that the flat bands induced by this form of dissipation in the Qi-Wu-Zhang model are topologically trivial.
    	    

\begin{thebibliography}{43}%
\makeatletter
\providecommand \@ifxundefined [1]{%
 \@ifx{#1\undefined}
}%
\providecommand \@ifnum [1]{%
 \ifnum #1\expandafter \@firstoftwo
 \else \expandafter \@secondoftwo
 \fi
}%
\providecommand \@ifx [1]{%
 \ifx #1\expandafter \@firstoftwo
 \else \expandafter \@secondoftwo
 \fi
}%
\providecommand \natexlab [1]{#1}%
\providecommand \enquote  [1]{``#1''}%
\providecommand \bibnamefont  [1]{#1}%
\providecommand \bibfnamefont [1]{#1}%
\providecommand \citenamefont [1]{#1}%
\providecommand \href@noop [0]{\@secondoftwo}%
\providecommand \href [0]{\begingroup \@sanitize@url \@href}%
\providecommand \@href[1]{\@@startlink{#1}\@@href}%
\providecommand \@@href[1]{\endgroup#1\@@endlink}%
\providecommand \@sanitize@url [0]{\catcode `\\12\catcode `\$12\catcode
  `\&12\catcode `\#12\catcode `\^12\catcode `\_12\catcode `\%12\relax}%
\providecommand \@@startlink[1]{}%
\providecommand \@@endlink[0]{}%
\providecommand \url  [0]{\begingroup\@sanitize@url \@url }%
\providecommand \@url [1]{\endgroup\@href {#1}{\urlprefix }}%
\providecommand \urlprefix  [0]{URL }%
\providecommand \Eprint [0]{\href }%
\providecommand \doibase [0]{https://doi.org/}%
\providecommand \selectlanguage [0]{\@gobble}%
\providecommand \bibinfo  [0]{\@secondoftwo}%
\providecommand \bibfield  [0]{\@secondoftwo}%
\providecommand \translation [1]{[#1]}%
\providecommand \BibitemOpen [0]{}%
\providecommand \bibitemStop [0]{}%
\providecommand \bibitemNoStop [0]{.\EOS\space}%
\providecommand \EOS [0]{\spacefactor3000\relax}%
\providecommand \BibitemShut  [1]{\csname bibitem#1\endcsname}%
\let\auto@bib@innerbib\@empty
\bibitem [{\citenamefont {Sy{\^o}zi}(1951)}]{syozi1951statistics}%
  \BibitemOpen
  \bibfield  {author} {\bibinfo {author} {\bibfnamefont {I.}~\bibnamefont
  {Sy{\^o}zi}},\ }\href {https://doi.org/10.1143/ptp/6.3.306} {\bibfield
  {journal} {\bibinfo  {journal} {Prog. Th. Phys.}\ }\textbf {\bibinfo {volume}
  {6}},\ \bibinfo {pages} {306} (\bibinfo {year} {1951})}\BibitemShut {NoStop}%
\bibitem [{\citenamefont {Lieb}(1989)}]{lieb1989two}%
  \BibitemOpen
  \bibfield  {author} {\bibinfo {author} {\bibfnamefont {E.~H.}\ \bibnamefont
  {Lieb}},\ }\href {https://doi.org/10.1103/PhysRevLett.62.1201} {\bibfield
  {journal} {\bibinfo  {journal} {Phys. Rev. Lett.}\ }\textbf {\bibinfo
  {volume} {62}},\ \bibinfo {pages} {1201} (\bibinfo {year}
  {1989})}\BibitemShut {NoStop}%
\bibitem [{\citenamefont {Suárez~Morell}\ \emph {et~al.}(2010)\citenamefont
  {Suárez~Morell}, \citenamefont {Correa}, \citenamefont {Vargas},
  \citenamefont {Pacheco},\ and\ \citenamefont
  {Barticevic}}]{suarez_morell_flat_2010}%
  \BibitemOpen
  \bibfield  {author} {\bibinfo {author} {\bibfnamefont {E.}~\bibnamefont
  {Suárez~Morell}}, \bibinfo {author} {\bibfnamefont {J.~D.}\ \bibnamefont
  {Correa}}, \bibinfo {author} {\bibfnamefont {P.}~\bibnamefont {Vargas}},
  \bibinfo {author} {\bibfnamefont {M.}~\bibnamefont {Pacheco}},\ and\ \bibinfo
  {author} {\bibfnamefont {Z.}~\bibnamefont {Barticevic}},\ }\href
  {https://doi.org/10.1103/PhysRevB.82.121407} {\bibfield  {journal} {\bibinfo
  {journal} {Phys. Rev. B}\ }\textbf {\bibinfo {volume} {82}},\ \bibinfo
  {pages} {121407} (\bibinfo {year} {2010})}\BibitemShut {NoStop}%
\bibitem [{\citenamefont {Bistritzer}\ and\ \citenamefont
  {MacDonald}(2011)}]{bistritzer_moire_2011}%
  \BibitemOpen
  \bibfield  {author} {\bibinfo {author} {\bibfnamefont {R.}~\bibnamefont
  {Bistritzer}}\ and\ \bibinfo {author} {\bibfnamefont {A.~H.}\ \bibnamefont
  {MacDonald}},\ }\href {https://doi.org/10.1073/pnas.1108174108} {\bibfield
  {journal} {\bibinfo  {journal} {Proc. Nat. Acad. Sci. U.S.A.}\ }\textbf
  {\bibinfo {volume} {108}},\ \bibinfo {pages} {12233} (\bibinfo {year}
  {2011})}\BibitemShut {NoStop}%
\bibitem [{\citenamefont {Regnault}\ \emph {et~al.}(2021)\citenamefont
  {Regnault}, \citenamefont {Xu}, \citenamefont {Li}, \citenamefont {Ma},
  \citenamefont {Jovanovic}, \citenamefont {Yazdani}, \citenamefont {Parkin},
  \citenamefont {Felser}, \citenamefont {Schoop},\ and\ \citenamefont
  {Ong}}]{regnault2021catalogue}%
  \BibitemOpen
  \bibfield  {author} {\bibinfo {author} {\bibfnamefont {N.}~\bibnamefont
  {Regnault}}, \bibinfo {author} {\bibfnamefont {Y.}~\bibnamefont {Xu}},
  \bibinfo {author} {\bibfnamefont {M.~R.}\ \bibnamefont {Li}}, \bibinfo
  {author} {\bibfnamefont {D.~S.}\ \bibnamefont {Ma}}, \bibinfo {author}
  {\bibfnamefont {M.}~\bibnamefont {Jovanovic}}, \bibinfo {author}
  {\bibfnamefont {A.}~\bibnamefont {Yazdani}}, \bibinfo {author} {\bibfnamefont
  {S.~S.}\ \bibnamefont {Parkin}}, \bibinfo {author} {\bibfnamefont
  {C.}~\bibnamefont {Felser}}, \bibinfo {author} {\bibfnamefont {L.~M.}\
  \bibnamefont {Schoop}},\ and\ \bibinfo {author} {\bibfnamefont {N.~P.}\
  \bibnamefont {Ong}},\ }\href {https://arxiv.org/abs/2106.05287} {\bibfield
  {journal} {\bibinfo  {journal} {arXiv:2106.05287}\ } (\bibinfo {year}
  {2021})}\BibitemShut {NoStop}%
\bibitem [{\citenamefont {C\u{a}lug\u{a}ru}\ \emph {et~al.}(2021)\citenamefont
  {C\u{a}lug\u{a}ru}, \citenamefont {Chew}, \citenamefont {Elcoro},
  \citenamefont {Regnault}, \citenamefont {Song},\ and\ \citenamefont
  {Bernevig}}]{calugaru21}%
  \BibitemOpen
  \bibfield  {author} {\bibinfo {author} {\bibfnamefont {D.}~\bibnamefont
  {C\u{a}lug\u{a}ru}}, \bibinfo {author} {\bibfnamefont {A.}~\bibnamefont
  {Chew}}, \bibinfo {author} {\bibfnamefont {L.}~\bibnamefont {Elcoro}},
  \bibinfo {author} {\bibfnamefont {N.}~\bibnamefont {Regnault}}, \bibinfo
  {author} {\bibfnamefont {Z.~D.}\ \bibnamefont {Song}},\ and\ \bibinfo
  {author} {\bibfnamefont {B.~A.}\ \bibnamefont {Bernevig}},\ }\href@noop {}
  {\bibfield  {journal} {\bibinfo  {journal} {Nature Phys.}\ } (\bibinfo {year}
  {2021})}\BibitemShut {NoStop}%
\bibitem [{\citenamefont {Julku}\ \emph {et~al.}(2016)\citenamefont {Julku},
  \citenamefont {Peotta}, \citenamefont {Vanhala}, \citenamefont {Kim},\ and\
  \citenamefont {T{\"o}rm{\"a}}}]{julku2016geometric}%
  \BibitemOpen
  \bibfield  {author} {\bibinfo {author} {\bibfnamefont {A.}~\bibnamefont
  {Julku}}, \bibinfo {author} {\bibfnamefont {S.}~\bibnamefont {Peotta}},
  \bibinfo {author} {\bibfnamefont {T.~I.}\ \bibnamefont {Vanhala}}, \bibinfo
  {author} {\bibfnamefont {D.-H.}\ \bibnamefont {Kim}},\ and\ \bibinfo {author}
  {\bibfnamefont {P.}~\bibnamefont {T{\"o}rm{\"a}}},\ }\href
  {https://doi.org/10.1103/PhysRevLett.117.045303} {\bibfield  {journal}
  {\bibinfo  {journal} {Phys. Rev. Lett.}\ }\textbf {\bibinfo {volume} {117}},\
  \bibinfo {pages} {045303} (\bibinfo {year} {2016})}\BibitemShut {NoStop}%
\bibitem [{\citenamefont {Liang}\ \emph {et~al.}(2017)\citenamefont {Liang},
  \citenamefont {Vanhala}, \citenamefont {Peotta}, \citenamefont {Siro},
  \citenamefont {Harju},\ and\ \citenamefont {T{\"o}rm{\"a}}}]{liang2017band}%
  \BibitemOpen
  \bibfield  {author} {\bibinfo {author} {\bibfnamefont {L.}~\bibnamefont
  {Liang}}, \bibinfo {author} {\bibfnamefont {T.~I.}\ \bibnamefont {Vanhala}},
  \bibinfo {author} {\bibfnamefont {S.}~\bibnamefont {Peotta}}, \bibinfo
  {author} {\bibfnamefont {T.}~\bibnamefont {Siro}}, \bibinfo {author}
  {\bibfnamefont {A.}~\bibnamefont {Harju}},\ and\ \bibinfo {author}
  {\bibfnamefont {P.}~\bibnamefont {T{\"o}rm{\"a}}},\ }\href
  {https://doi.org/10.1103/PhysRevB.95.024515} {\bibfield  {journal} {\bibinfo
  {journal} {Phys. Rev. B}\ }\textbf {\bibinfo {volume} {95}},\ \bibinfo
  {pages} {024515} (\bibinfo {year} {2017})}\BibitemShut {NoStop}%
\bibitem [{\citenamefont {Herzog-Arbeitman}\ \emph {et~al.}(2022)\citenamefont
  {Herzog-Arbeitman}, \citenamefont {Peri}, \citenamefont {Schindler},
  \citenamefont {Huber},\ and\ \citenamefont
  {Bernevig}}]{herzog2022superfluid}%
  \BibitemOpen
  \bibfield  {author} {\bibinfo {author} {\bibfnamefont {J.}~\bibnamefont
  {Herzog-Arbeitman}}, \bibinfo {author} {\bibfnamefont {V.}~\bibnamefont
  {Peri}}, \bibinfo {author} {\bibfnamefont {F.}~\bibnamefont {Schindler}},
  \bibinfo {author} {\bibfnamefont {S.~D.}\ \bibnamefont {Huber}},\ and\
  \bibinfo {author} {\bibfnamefont {B.~A.}\ \bibnamefont {Bernevig}},\ }\href
  {https://doi.org/10.1103/PhysRevLett.128.087002} {\bibfield  {journal}
  {\bibinfo  {journal} {Phys. Rev. Lett.}\ }\textbf {\bibinfo {volume} {128}},\
  \bibinfo {pages} {087002} (\bibinfo {year} {2022})}\BibitemShut {NoStop}%
\bibitem [{\citenamefont {Bauer}\ \emph {et~al.}(2022)\citenamefont {Bauer},
  \citenamefont {Talkington}, \citenamefont {Harper}, \citenamefont {Andrews},\
  and\ \citenamefont {Roy}}]{bauer2022fractional}%
  \BibitemOpen
  \bibfield  {author} {\bibinfo {author} {\bibfnamefont {D.}~\bibnamefont
  {Bauer}}, \bibinfo {author} {\bibfnamefont {S.}~\bibnamefont {Talkington}},
  \bibinfo {author} {\bibfnamefont {F.}~\bibnamefont {Harper}}, \bibinfo
  {author} {\bibfnamefont {B.}~\bibnamefont {Andrews}},\ and\ \bibinfo {author}
  {\bibfnamefont {R.}~\bibnamefont {Roy}},\ }\href
  {https://doi.org/10.1103/PhysRevB.105.045144} {\bibfield  {journal} {\bibinfo
   {journal} {Phys. Rev. B}\ }\textbf {\bibinfo {volume} {105}},\ \bibinfo
  {pages} {045144} (\bibinfo {year} {2022})}\BibitemShut {NoStop}%
\bibitem [{\citenamefont {Abouelkomsan}\ \emph {et~al.}(2022)\citenamefont
  {Abouelkomsan}, \citenamefont {Yang},\ and\ \citenamefont
  {Bergholtz}}]{abouelkomsan2022quantum}%
  \BibitemOpen
  \bibfield  {author} {\bibinfo {author} {\bibfnamefont {A.}~\bibnamefont
  {Abouelkomsan}}, \bibinfo {author} {\bibfnamefont {K.}~\bibnamefont {Yang}},\
  and\ \bibinfo {author} {\bibfnamefont {E.~J.}\ \bibnamefont {Bergholtz}},\
  }\bibfield  {journal} {\bibinfo  {journal} {arXiv 2202.10467}\ }\href
  {https://doi.org/10.48550/arXiv.2202.10467} {10.48550/arXiv.2202.10467}
  (\bibinfo {year} {2022})\BibitemShut {NoStop}%
\bibitem [{\citenamefont {Wang}\ \emph {et~al.}(2020)\citenamefont {Wang},
  \citenamefont {Shih}, \citenamefont {Ghiotto}, \citenamefont {Xian},
  \citenamefont {Rhodes}, \citenamefont {Tan}, \citenamefont {Claassen},
  \citenamefont {Kennes}, \citenamefont {Bai},\ and\ \citenamefont
  {Kim}}]{wang2020correlated}%
  \BibitemOpen
  \bibfield  {author} {\bibinfo {author} {\bibfnamefont {L.}~\bibnamefont
  {Wang}}, \bibinfo {author} {\bibfnamefont {E.-M.}\ \bibnamefont {Shih}},
  \bibinfo {author} {\bibfnamefont {A.}~\bibnamefont {Ghiotto}}, \bibinfo
  {author} {\bibfnamefont {L.}~\bibnamefont {Xian}}, \bibinfo {author}
  {\bibfnamefont {D.~A.}\ \bibnamefont {Rhodes}}, \bibinfo {author}
  {\bibfnamefont {C.}~\bibnamefont {Tan}}, \bibinfo {author} {\bibfnamefont
  {M.}~\bibnamefont {Claassen}}, \bibinfo {author} {\bibfnamefont {D.~M.}\
  \bibnamefont {Kennes}}, \bibinfo {author} {\bibfnamefont {Y.}~\bibnamefont
  {Bai}},\ and\ \bibinfo {author} {\bibfnamefont {B.}~\bibnamefont {Kim}},\
  }\href {https://doi.org/https://doi.org/10.1038/s41563-020-0708-6} {\bibfield
   {journal} {\bibinfo  {journal} {Nature Mater.}\ }\textbf {\bibinfo {volume}
  {19}},\ \bibinfo {pages} {861} (\bibinfo {year} {2020})}\BibitemShut
  {NoStop}%
\bibitem [{\citenamefont {Devakul}\ \emph {et~al.}(2021)\citenamefont
  {Devakul}, \citenamefont {Cr\'epel}, \citenamefont {Zhang},\ and\
  \citenamefont {Fu}}]{devakul2021}%
  \BibitemOpen
  \bibfield  {author} {\bibinfo {author} {\bibfnamefont {T.}~\bibnamefont
  {Devakul}}, \bibinfo {author} {\bibfnamefont {V.}~\bibnamefont {Cr\'epel}},
  \bibinfo {author} {\bibfnamefont {Y.}~\bibnamefont {Zhang}},\ and\ \bibinfo
  {author} {\bibfnamefont {L.}~\bibnamefont {Fu}},\ }\href@noop {} {\bibfield
  {journal} {\bibinfo  {journal} {Nature Comm.}\ }\textbf {\bibinfo {volume}
  {12}},\ \bibinfo {pages} {6730} (\bibinfo {year} {2021})}\BibitemShut
  {NoStop}%
\bibitem [{\citenamefont {Diehl}\ \emph {et~al.}(2011)\citenamefont {Diehl},
  \citenamefont {Rico}, \citenamefont {Baranov},\ and\ \citenamefont
  {Zoller}}]{diehl2011topology}%
  \BibitemOpen
  \bibfield  {author} {\bibinfo {author} {\bibfnamefont {S.}~\bibnamefont
  {Diehl}}, \bibinfo {author} {\bibfnamefont {E.}~\bibnamefont {Rico}},
  \bibinfo {author} {\bibfnamefont {M.~A.}\ \bibnamefont {Baranov}},\ and\
  \bibinfo {author} {\bibfnamefont {P.}~\bibnamefont {Zoller}},\ }\href
  {https://doi.org/10.1038/nphys2106} {\bibfield  {journal} {\bibinfo
  {journal} {Nat. Phys.}\ }\textbf {\bibinfo {volume} {7}},\ \bibinfo {pages}
  {971} (\bibinfo {year} {2011})}\BibitemShut {NoStop}%
\bibitem [{\citenamefont {Bardyn}\ \emph {et~al.}(2013)\citenamefont {Bardyn},
  \citenamefont {Baranov}, \citenamefont {Kraus}, \citenamefont {Rico},
  \citenamefont {{\.I}mamo{\u{g}}lu}, \citenamefont {Zoller},\ and\
  \citenamefont {Diehl}}]{bardyn2013topology}%
  \BibitemOpen
  \bibfield  {author} {\bibinfo {author} {\bibfnamefont {C.~E.}\ \bibnamefont
  {Bardyn}}, \bibinfo {author} {\bibfnamefont {M.~A.}\ \bibnamefont {Baranov}},
  \bibinfo {author} {\bibfnamefont {C.~V.}\ \bibnamefont {Kraus}}, \bibinfo
  {author} {\bibfnamefont {E.}~\bibnamefont {Rico}}, \bibinfo {author}
  {\bibfnamefont {A.}~\bibnamefont {{\.I}mamo{\u{g}}lu}}, \bibinfo {author}
  {\bibfnamefont {P.}~\bibnamefont {Zoller}},\ and\ \bibinfo {author}
  {\bibfnamefont {S.}~\bibnamefont {Diehl}},\ }\href
  {https://doi.org/10.1088/1367-2630/15/8/085001} {\bibfield  {journal}
  {\bibinfo  {journal} {New J. Phys.}\ }\textbf {\bibinfo {volume} {15}},\
  \bibinfo {pages} {085001} (\bibinfo {year} {2013})}\BibitemShut {NoStop}%
\bibitem [{\citenamefont {Goldstein}(2019)}]{goldstein2019dissipation}%
  \BibitemOpen
  \bibfield  {author} {\bibinfo {author} {\bibfnamefont {M.}~\bibnamefont
  {Goldstein}},\ }\href {https://doi.org/10.21468/SciPostPhys.7.5.067}
  {\bibfield  {journal} {\bibinfo  {journal} {SciPost Phys.}\ }\textbf
  {\bibinfo {volume} {7}},\ \bibinfo {pages} {67} (\bibinfo {year}
  {2019})}\BibitemShut {NoStop}%
\bibitem [{\citenamefont {Huang}\ and\ \citenamefont
  {Arovas}(2014)}]{huang2014topological}%
  \BibitemOpen
  \bibfield  {author} {\bibinfo {author} {\bibfnamefont {Z.}~\bibnamefont
  {Huang}}\ and\ \bibinfo {author} {\bibfnamefont {D.~P.}\ \bibnamefont
  {Arovas}},\ }\href {https://doi.org/10.1103/PhysRevLett.113.076407}
  {\bibfield  {journal} {\bibinfo  {journal} {Phys. Rev. Lett.}\ }\textbf
  {\bibinfo {volume} {113}},\ \bibinfo {pages} {076407} (\bibinfo {year}
  {2014})}\BibitemShut {NoStop}%
\bibitem [{\citenamefont {Lieu}(2018)}]{lieu2018topological}%
  \BibitemOpen
  \bibfield  {author} {\bibinfo {author} {\bibfnamefont {S.}~\bibnamefont
  {Lieu}},\ }\href {https://doi.org/10.1103/PhysRevB.98.115135} {\bibfield
  {journal} {\bibinfo  {journal} {Phys. Rev. B}\ }\textbf {\bibinfo {volume}
  {98}},\ \bibinfo {pages} {115135} (\bibinfo {year} {2018})}\BibitemShut
  {NoStop}%
\bibitem [{\citenamefont {Gong}\ \emph {et~al.}(2018)\citenamefont {Gong},
  \citenamefont {Ashida}, \citenamefont {Kawabata}, \citenamefont {Takasan},
  \citenamefont {Higashikawa},\ and\ \citenamefont
  {Ueda}}]{gong2018topological}%
  \BibitemOpen
  \bibfield  {author} {\bibinfo {author} {\bibfnamefont {Z.}~\bibnamefont
  {Gong}}, \bibinfo {author} {\bibfnamefont {Y.}~\bibnamefont {Ashida}},
  \bibinfo {author} {\bibfnamefont {K.}~\bibnamefont {Kawabata}}, \bibinfo
  {author} {\bibfnamefont {K.}~\bibnamefont {Takasan}}, \bibinfo {author}
  {\bibfnamefont {S.}~\bibnamefont {Higashikawa}},\ and\ \bibinfo {author}
  {\bibfnamefont {M.}~\bibnamefont {Ueda}},\ }\href
  {https://doi.org/10.1103/PhysRevX.8.031079} {\bibfield  {journal} {\bibinfo
  {journal} {Phys. Rev. X}\ }\textbf {\bibinfo {volume} {8}},\ \bibinfo {pages}
  {031079} (\bibinfo {year} {2018})}\BibitemShut {NoStop}%
\bibitem [{\citenamefont {Zhang}\ and\ \citenamefont
  {Gong}(2018)}]{zhang2018topological}%
  \BibitemOpen
  \bibfield  {author} {\bibinfo {author} {\bibfnamefont {D.~J.}\ \bibnamefont
  {Zhang}}\ and\ \bibinfo {author} {\bibfnamefont {J.}~\bibnamefont {Gong}},\
  }\href {https://doi.org/10.1103/PhysRevA.98.052101} {\bibfield  {journal}
  {\bibinfo  {journal} {Phys. Rev. A}\ }\textbf {\bibinfo {volume} {98}},\
  \bibinfo {pages} {052101} (\bibinfo {year} {2018})}\BibitemShut {NoStop}%
\bibitem [{\citenamefont {Ramezani}(2017)}]{ramezani2017non}%
  \BibitemOpen
  \bibfield  {author} {\bibinfo {author} {\bibfnamefont {H.}~\bibnamefont
  {Ramezani}},\ }\href {https://doi.org/10.1103/PhysRevA.96.011802} {\bibfield
  {journal} {\bibinfo  {journal} {Phys. Rev. A}\ }\textbf {\bibinfo {volume}
  {96}},\ \bibinfo {pages} {011802} (\bibinfo {year} {2017})}\BibitemShut
  {NoStop}%
\bibitem [{\citenamefont {Ge}(2018)}]{ge2018non}%
  \BibitemOpen
  \bibfield  {author} {\bibinfo {author} {\bibfnamefont {L.}~\bibnamefont
  {Ge}},\ }\href {https://doi.org/10.1364/PRJ.6.000A10} {\bibfield  {journal}
  {\bibinfo  {journal} {Photon. Res.}\ }\textbf {\bibinfo {volume} {6}},\
  \bibinfo {pages} {A10} (\bibinfo {year} {2018})}\BibitemShut {NoStop}%
\bibitem [{\citenamefont {Lindblad}(1976)}]{lindblad1976generators}%
  \BibitemOpen
  \bibfield  {author} {\bibinfo {author} {\bibfnamefont {G.}~\bibnamefont
  {Lindblad}},\ }\href {https://doi.org/10.1007/BF01608499} {\bibfield
  {journal} {\bibinfo  {journal} {Comm. Math. Phys.}\ }\textbf {\bibinfo
  {volume} {48}},\ \bibinfo {pages} {119} (\bibinfo {year} {1976})}\BibitemShut
  {NoStop}%
\bibitem [{\citenamefont {Gorini}\ \emph {et~al.}(1976)\citenamefont {Gorini},
  \citenamefont {Kossakowski},\ and\ \citenamefont
  {Sudarshan}}]{gorini1976completely}%
  \BibitemOpen
  \bibfield  {author} {\bibinfo {author} {\bibfnamefont {V.}~\bibnamefont
  {Gorini}}, \bibinfo {author} {\bibfnamefont {A.}~\bibnamefont
  {Kossakowski}},\ and\ \bibinfo {author} {\bibfnamefont {E.~C.~G.}\
  \bibnamefont {Sudarshan}},\ }\href {https://doi.org/10.1063/1.522979}
  {\bibfield  {journal} {\bibinfo  {journal} {J. Math. Phys.}\ }\textbf
  {\bibinfo {volume} {17}},\ \bibinfo {pages} {821} (\bibinfo {year}
  {1976})}\BibitemShut {NoStop}%
\bibitem [{\citenamefont {Gardiner}\ \emph {et~al.}(2004)\citenamefont
  {Gardiner}, \citenamefont {Zoller},\ and\ \citenamefont
  {Zoller}}]{gardiner2004quantum}%
  \BibitemOpen
  \bibfield  {author} {\bibinfo {author} {\bibfnamefont {C.}~\bibnamefont
  {Gardiner}}, \bibinfo {author} {\bibfnamefont {P.}~\bibnamefont {Zoller}},\
  and\ \bibinfo {author} {\bibfnamefont {P.}~\bibnamefont {Zoller}},\ }\href
  {https://link.springer.com/book/9783540223016} {\emph {\bibinfo {title}
  {{Quantum Noise}}}}\ (\bibinfo  {publisher} {Springer Science},\ \bibinfo
  {year} {2004})\BibitemShut {NoStop}%
\bibitem [{\citenamefont {Am-Shallem}\ \emph {et~al.}(2015)\citenamefont
  {Am-Shallem}, \citenamefont {Levy}, \citenamefont {Schaefer},\ and\
  \citenamefont {Kosloff}}]{amshallem2015three}%
  \BibitemOpen
  \bibfield  {author} {\bibinfo {author} {\bibfnamefont {M.}~\bibnamefont
  {Am-Shallem}}, \bibinfo {author} {\bibfnamefont {A.}~\bibnamefont {Levy}},
  \bibinfo {author} {\bibfnamefont {I.}~\bibnamefont {Schaefer}},\ and\
  \bibinfo {author} {\bibfnamefont {R.}~\bibnamefont {Kosloff}},\ }\href
  {https://arxiv.org/abs/1510.08634} {\bibfield  {journal} {\bibinfo  {journal}
  {arXiv:1510.08634}\ } (\bibinfo {year} {2015})}\BibitemShut {NoStop}%
\bibitem [{\citenamefont {Manzano}(2020)}]{manzano2020short}%
  \BibitemOpen
  \bibfield  {author} {\bibinfo {author} {\bibfnamefont {D.}~\bibnamefont
  {Manzano}},\ }\href {https://doi.org/10.1063/1.5115323} {\bibfield  {journal}
  {\bibinfo  {journal} {AIP Advances}\ }\textbf {\bibinfo {volume} {10}},\
  \bibinfo {pages} {025106} (\bibinfo {year} {2020})}\BibitemShut {NoStop}%
\bibitem [{\citenamefont {Altland}\ and\ \citenamefont
  {Simons}(2010)}]{altland2010condensed}%
  \BibitemOpen
  \bibfield  {author} {\bibinfo {author} {\bibfnamefont {A.}~\bibnamefont
  {Altland}}\ and\ \bibinfo {author} {\bibfnamefont {B.~D.}\ \bibnamefont
  {Simons}},\ }\href
  {https://www.cambridge.org/core/books/condensed-matter-field-theory/0A8DE6503ED868D96985D9E7847C63FF}
  {\emph {\bibinfo {title} {Condensed Matter Field Theory}}}\ (\bibinfo
  {publisher} {Cambridge University Press},\ \bibinfo {year}
  {2010})\BibitemShut {NoStop}%
\bibitem [{\citenamefont {Walls}(1983)}]{walls1983squeezed}%
  \BibitemOpen
  \bibfield  {author} {\bibinfo {author} {\bibfnamefont {D.~F.}\ \bibnamefont
  {Walls}},\ }\href {https://doi.org/10.1038/306141a0} {\bibfield  {journal}
  {\bibinfo  {journal} {Nature}\ }\textbf {\bibinfo {volume} {306}},\ \bibinfo
  {pages} {141} (\bibinfo {year} {1983})}\BibitemShut {NoStop}%
\bibitem [{\citenamefont {Vaccaro}\ and\ \citenamefont
  {Pegg}(1989)}]{vaccaro1989phase}%
  \BibitemOpen
  \bibfield  {author} {\bibinfo {author} {\bibfnamefont {J.}~\bibnamefont
  {Vaccaro}}\ and\ \bibinfo {author} {\bibfnamefont {D.}~\bibnamefont {Pegg}},\
  }\href {https://doi.org/10.1016/0030-4018(89)90377-5} {\bibfield  {journal}
  {\bibinfo  {journal} {Optics Commun.}\ }\textbf {\bibinfo {volume} {70}},\
  \bibinfo {pages} {529} (\bibinfo {year} {1989})}\BibitemShut {NoStop}%
\bibitem [{\citenamefont {Prosen}(2008)}]{prosen2008third}%
  \BibitemOpen
  \bibfield  {author} {\bibinfo {author} {\bibfnamefont {T.}~\bibnamefont
  {Prosen}},\ }\href {https://doi.org/10.1088/1367-2630/10/4/043026} {\bibfield
   {journal} {\bibinfo  {journal} {New J. Phys.}\ }\textbf {\bibinfo {volume}
  {10}},\ \bibinfo {pages} {043026} (\bibinfo {year} {2008})}\BibitemShut
  {NoStop}%
\bibitem [{\citenamefont {Prosen}(2010)}]{prosen2010spectral}%
  \BibitemOpen
  \bibfield  {author} {\bibinfo {author} {\bibfnamefont {T.}~\bibnamefont
  {Prosen}},\ }\href {https://doi.org/10.1088/1742-5468/2010/07/P07020}
  {\bibfield  {journal} {\bibinfo  {journal} {J. Stat. Mech.}\ }\textbf
  {\bibinfo {volume} {2010}},\ \bibinfo {pages} {P07020} (\bibinfo {year}
  {2010})}\BibitemShut {NoStop}%
\bibitem [{\citenamefont {Lieu}\ \emph {et~al.}(2020)\citenamefont {Lieu},
  \citenamefont {McGinley},\ and\ \citenamefont {Cooper}}]{lieu2020tenfold}%
  \BibitemOpen
  \bibfield  {author} {\bibinfo {author} {\bibfnamefont {S.}~\bibnamefont
  {Lieu}}, \bibinfo {author} {\bibfnamefont {M.}~\bibnamefont {McGinley}},\
  and\ \bibinfo {author} {\bibfnamefont {N.~R.}\ \bibnamefont {Cooper}},\
  }\href {https://doi.org/10.1103/PhysRevLett.124.040401} {\bibfield  {journal}
  {\bibinfo  {journal} {Phys. Rev. Lett.}\ }\textbf {\bibinfo {volume} {124}},\
  \bibinfo {pages} {040401} (\bibinfo {year} {2020})}\BibitemShut {NoStop}%
\bibitem [{\citenamefont {Sayyad}\ \emph {et~al.}(2021)\citenamefont {Sayyad},
  \citenamefont {Yu}, \citenamefont {Grushin},\ and\ \citenamefont
  {Sieberer}}]{sayyad2021entanglement}%
  \BibitemOpen
  \bibfield  {author} {\bibinfo {author} {\bibfnamefont {S.}~\bibnamefont
  {Sayyad}}, \bibinfo {author} {\bibfnamefont {J.}~\bibnamefont {Yu}}, \bibinfo
  {author} {\bibfnamefont {A.~G.}\ \bibnamefont {Grushin}},\ and\ \bibinfo
  {author} {\bibfnamefont {L.~M.}\ \bibnamefont {Sieberer}},\ }\href
  {https://doi.org/10.1103/PhysRevResearch.3.033022} {\bibfield  {journal}
  {\bibinfo  {journal} {Phys. Rev. Res.}\ }\textbf {\bibinfo {volume} {3}},\
  \bibinfo {pages} {033022} (\bibinfo {year} {2021})}\BibitemShut {NoStop}%
\bibitem [{\citenamefont {Shavit}\ and\ \citenamefont
  {Goldstein}(2020)}]{shavit2020topology}%
  \BibitemOpen
  \bibfield  {author} {\bibinfo {author} {\bibfnamefont {G.}~\bibnamefont
  {Shavit}}\ and\ \bibinfo {author} {\bibfnamefont {M.}~\bibnamefont
  {Goldstein}},\ }\href {https://doi.org/10.1103/PhysRevB.101.125412}
  {\bibfield  {journal} {\bibinfo  {journal} {Phys. Rev. B}\ }\textbf {\bibinfo
  {volume} {101}},\ \bibinfo {pages} {125412} (\bibinfo {year}
  {2020})}\BibitemShut {NoStop}%
\bibitem [{\citenamefont {Beck}\ and\ \citenamefont
  {Goldstein}(2021)}]{beck2021disorder}%
  \BibitemOpen
  \bibfield  {author} {\bibinfo {author} {\bibfnamefont {A.}~\bibnamefont
  {Beck}}\ and\ \bibinfo {author} {\bibfnamefont {M.}~\bibnamefont
  {Goldstein}},\ }\href {https://doi.org/10.1103/PhysRevB.103.L241401}
  {\bibfield  {journal} {\bibinfo  {journal} {Phys. Rev. B}\ }\textbf {\bibinfo
  {volume} {103}},\ \bibinfo {pages} {L241401} (\bibinfo {year}
  {2021})}\BibitemShut {NoStop}%
\bibitem [{Note1()}]{Note1}%
  \BibitemOpen
  \bibinfo {note} {The transposition for charge conjugation, as opposed to
  complex conjugation, stems from $L$ being non-Hermitian \cite
  {gong2018topological}.}\BibitemShut {Stop}%
\bibitem [{Note2()}]{Note2}%
  \BibitemOpen
  \bibinfo {note} {Here we show the case where $H:=H_{\protect \bm
  {k}}=H_{-\protect \bm {k}}$. For the general case, see the [Supplemental
  Material].}\BibitemShut {Stop}%
\bibitem [{Note3()}]{Note3}%
  \BibitemOpen
  \bibinfo {note} {We conjecture that this is the \protect \textit {only} form
  that ensures $\protect \mathcal {D}$ symmetry.}\BibitemShut {Stop}%
\bibitem [{\citenamefont {Hatsugai}\ and\ \citenamefont
  {Maruyama}(2011)}]{hatsugai2011zq}%
  \BibitemOpen
  \bibfield  {author} {\bibinfo {author} {\bibfnamefont {Y.}~\bibnamefont
  {Hatsugai}}\ and\ \bibinfo {author} {\bibfnamefont {I.}~\bibnamefont
  {Maruyama}},\ }\href {https://doi.org/10.1209/0295-5075/95/20003} {\bibfield
  {journal} {\bibinfo  {journal} {EPL}\ }\textbf {\bibinfo {volume} {95}},\
  \bibinfo {pages} {20003} (\bibinfo {year} {2011})}\BibitemShut {NoStop}%
\bibitem [{\citenamefont {Mizoguchi}\ and\ \citenamefont
  {Hatsugai}(2019)}]{mizoguchi2019molecular}%
  \BibitemOpen
  \bibfield  {author} {\bibinfo {author} {\bibfnamefont {T.}~\bibnamefont
  {Mizoguchi}}\ and\ \bibinfo {author} {\bibfnamefont {Y.}~\bibnamefont
  {Hatsugai}},\ }\href {https://doi.org/10.1209/0295-5075/127/47001} {\bibfield
   {journal} {\bibinfo  {journal} {EPL}\ }\textbf {\bibinfo {volume} {127}},\
  \bibinfo {pages} {47001} (\bibinfo {year} {2019})}\BibitemShut {NoStop}%
\bibitem [{Note4()}]{Note4}%
  \BibitemOpen
  \bibinfo {note} {Here, factors proportional to the electronic bandwidth have
  been omitted for brevity.}\BibitemShut {Stop}%
\bibitem [{\citenamefont {Qi}\ \emph {et~al.}(2006)\citenamefont {Qi},
  \citenamefont {Wu},\ and\ \citenamefont {Zhang}}]{qi2006topological}%
  \BibitemOpen
  \bibfield  {author} {\bibinfo {author} {\bibfnamefont {X.~L.}\ \bibnamefont
  {Qi}}, \bibinfo {author} {\bibfnamefont {Y.~S.}\ \bibnamefont {Wu}},\ and\
  \bibinfo {author} {\bibfnamefont {S.~C.}\ \bibnamefont {Zhang}},\ }\href
  {https://doi.org/10.1103/PhysRevB.74.085308} {\bibfield  {journal} {\bibinfo
  {journal} {Phys. Rev. B}\ }\textbf {\bibinfo {volume} {74}},\ \bibinfo
  {pages} {085308} (\bibinfo {year} {2006})}\BibitemShut {NoStop}%
\end{thebibliography}

\end{document}